\documentclass[12pt]{iopart}
\usepackage[dvips]{graphicx}
\usepackage{amssymb}
\usepackage{latexsym}

\begin{document}
\frenchspacing

\def\picbox#1#2{\fbox{\vbox to#2{\hbox to#1{}}}}
\def\bra#1{\langle#1|}
\def\ket#1{|#1\rangle}
\def\braket#1#2{\langle#1|#2\rangle}
\def\ave#1{\left\langle #1 \right\rangle}
\def\parc#1#2{\frac{\partial #1}{\partial #2}}
\def\rot{\textrm{rot}}
\def\grad{\textrm{grad}}
\def\pa{\partial}
\def\scalar#1#2{\langle#1|#2\rangle}
\def\diag{{\rm diag}}
\def\card{{\rm card}}
\def\const{{\rm const}}
\def\eps{\varepsilon}
\def\bJ{\!\!\stackrel{\hskip1mm\leftrightarrow}{J}}
\def\mean#1{\overline{#1}}
\def \mod{~{\rm mod}~}
\def\C{{\mathbb{C}}}
\def\Z{{\mathbb{Z}}}
\def\R{{\mathbb{R}}}
\def\T{{\mathbb{T}}}
\def\N{{\mathbb{N}}}

\def\dd{{\rm d}}
\def\ii{{\rm i}}
\def\iint{\int\!\!\!\int}
\def\id{{\rm id}}

\def\asin{{\rm arcsin}}
\def\acos{{\rm arccos}}
\def\asinh{{\rm arcsinh}}
\def\acosh{{\rm arccosh}}
\def\atan{{\rm arctan}}

\newcommand{\ps}{X}

\newcommand{\erf}{\mathop{\rm erf}}
\newcommand{\erfc}{\mathop{\rm erfc}}
\newcommand{\erfi}{\mathop{\rm erf{}i}}
\newcommand{\sgn}{\mathop{\rm sign}}

\newcommand{\hash}{\#}
\newcommand{\eqcolon}{=:}
\newcommand{\coloneq}{:=}


\newcommand{\textmod}{\;{\mbox{\rm mod}}\:}
\newcommand{\I}{{1\!\!1}}
\newcommand{\1}{1\!\!1}
\newcommand{\hrho}{{\hat\rho}}
\newcommand{\hT}{{\hat T}}
\newcommand{\hq}{{\hat q}}
\newcommand{\hp}{{\hat p}}
\newcommand{\blds}{\boldsymbol}
\newcommand{\IR}{\mathbb{R}}
\newcommand{\IZ}{\mathbb{Z}}
\newcommand{\bv}{{\mathbf v}}
\newcommand{\ml}{\mathcal} 
\newcommand{\cH}{\ml{H}}
\newcommand{\bequ}{\begin{equation}}
\newcommand{\eequ}{\end{equation}}
\def\e{{\rm e}}
\newcommand{\cS}{\ml{S}}
\newcommand{\IN}{\mathbb{N}}
\def\t2{{\mathbb T}^2}
\newcommand{\bm}{\mathbf{m}}
\def\hn{\mathcal{H}_{N}}
\newcommand{\bq}{{q}}
\newcommand{\bp}{{p}}
\newcommand{\lt}{<}
\newcommand{\Cy}{\mathcal{C}}

\newcommand{\la}{\langle}
\newcommand{\IC}{\mathbb{C}} 
\newcommand{\IT}{\mathbb{T}} 

\newcommand{\ra}{\rangle}
\newcommand{\Op}[1]{{\mbox{Op}_W\{#1\}}}
\newcommand{\Opinv}[1]{{\mbox{Op}_W^{-1}\{#1\}}}
\newcommand{\eqref}[1]{(\ref{#1})}
\newcommand{\bk}{{\bf k}}
\newcommand{\bfm}{{\bf m}}
\newcommand{\bl}{{\bf \ell}}


\title{Quantum-classical correspondence on compact phase space}

\author{Martin Horvat$^{(1)}$,
        Toma\v z Prosen$^{(1)}$,
    Mirko Degli Esposti$^{(2)}$}

\address{
  ${}^{(1)}$ Physics Department, Faculty of Mathematics and Physics,
University of Ljubljana, Slovenia\\
  ${}^{(2)}$ Department of Mathematics, University of Bologna, Italy
}

\eads{\mailto{martin.horvat@fmf.uni-lj.si},
      \mailto{tomaz.prosen@fmf.uni-lj.si},
      \mailto{desposti@dm.unibo.it}}
\begin{abstract}
We propose to study the $L^2$-norm distance between classical and
quantum phase space distributions, where for the latter we choose
the Wigner function, as a global phase space indicator of
quantum-classical correspondence. For example, this quantity
should provide a key to understand the correspondence between
quantum and classical Loschmidt echoes. We concentrate on fully
chaotic systems with compact (finite) classical phase space. By
means of numerical simulations and heuristic arguments we find
that the quantum-classical fidelity stays at one up to
Ehrenfest-type time scale, which is proportional to the logarithm
of effective Planck constant, and decays exponentially with a
maximal classical Lyapunov exponent, after that time.
\end{abstract}
\submitto{Nonlinearity
}
\section{Introduction}
In this paper we want to address few questions concerning the correspondence of quantum and classical evolution on the classical phase space for quantum pure states of classically (strongly and weakly) chaotic systems. Our description is restricted to the compact or effectively compact phase space. By effectively, we mean that the classically available space is compact and quantum description can be effectively done in some finite dimensional set of basis functions.\par
We will use the Weyl-Wigner phase-space representations of quantum mechanical states \cite{lee95} and in this way compare the classical and the quantum evolution, both as evolution of functions defined over the classical phase space.
The analysis will be performed by considering coherent packets as initial quantum states, such that the corresponding Wigner functions coincides with the initial classical phase space distributions identical to Gaussian packets. It is a common knowledge in the realm of quantum chaos that the phase space correspondence between classical and quantum mechanics drops down on the scale of Ehrenfest time $t_{\rm E} = -\log \hbar/\lambda$, where $\hbar$ is an effective Planck constant $\hbar = \hbar_{\rm physical}/A$, $A$ is a typical action of the system and $\lambda$ is the maximal Lyapunov exponent. For a compact phase space the Ehrenfest time is usually defined as $t_{\rm E} =\log N/\lambda$, where $N$ is the Hilbert space dimension. Here we would like to investigate in more detail the break down of the correspondence and the corresponding break time scale.\par
Let us systematically define the problem. First of all, we start
with a classical Hamiltonian system defined over a given (i.e.
sympletic ) phase space $X$, for example $X=T^*\mathbb{R}^n$ or
$X=\T^2$. The trajectory of this system at the continuous
(discrete) time $t\in R^+$ ($t\in\mathbb{N}\cup\{0\}$), initially
in $x\in X$, is given by $\phi^t(x)\in X$. The corresponding
quantum system is defined over the Hilbert space $\cal H$ and is
described by a state $\ket{\psi}\in {\cal H}$. The motion of the
quantum state is given by the unitary evolution operator $U^t$,
with $t\in R^+$ and $t\in\mathbb{N}\cup\{0\}$, respectively.\par
To represent the quantum operators, $\hat A:{\cal H} \to {\cal H}$, as functions over the phase space $X$ we use the Wigner representation $\hat A\to W[\hat A](x)$ which share the same main properties in all classical phase-spaces, even if its specific form strongly depends on the geometry of $X$. We will recall the exact definition of these quantities on Euclidean $X=\mathbb{R}^{2d}$ and toroidal phase space $X=\mathbb{T}^2$. We are only interested in the phase space representation of density operators of pure states $\hat\rho = \ket{\psi}\bra{\psi}$ called the Wigner functions and denoted by $W_\psi(x) = W[\hat\rho](x)\in \mathbb{R}$ and normalized as $\int_X \dd x\,W_\psi(x) = 1$. By using the Wigner function one can represent quantum states and their dynamics in terms of phase-space functions. Similarly we can describe the state of a classical system on the phase space in terms of probability density functions \cite{arnold89}. \par
We study the {\it classical-quantum correspondence} (CQC) by
observing the deviation between the Wigner function of the quantum
system and the corresponding probability density of the classical
system evolving in time. The initial state $\ket{\psi(0)}$ of the
quantum system is a coherent state of minimal width with the
Wigner function $W^{t=0}(x)$ depending on the geometry of $X$, but
is basically similar to a symmetric Gaussian probability
distribution\footnote{On compact phase spaces it is sometimes
necessary to periodicize the Gaussian distribution to obtain
maximal similarity with the Wigner function of a coherent state.}
or is converging to its form in semi-classical limit $\hbar\to 0$
with error approximately of order $\exp(-|O(\hbar^{-1})|)$. The
last estimate is refined in the following sections for the toroidal
geometry  using the reference \cite{degliesposti05}. The state of
a classical system at $t=0$, denoted by $\rho^{t=0}(x)$, is a
Gaussian probability distribution corresponding to $W^{t=0}(x)$.
The phase-space functions of the initial quantum and classical
states are either identical, or converge to each other in the
semi-classical limit $\hbar\to 0$. Our aim is to understand the
behavior of the $L^2$ norm of the difference between the time
evolving Wigner function of the quantum system $W^t(x) = W_{U^t
\ket{\psi(0)}}(x)$ and the time evolving probability density of
the classical system $\rho^t (x) = \rho^0(\phi^{-t}(x))$, defined
as
\begin{equation}
  \| \rho^t - W^t \|_2^2 =
  \|\rho^t(x)\|_2^2  + \|W^t(x)\|_2^2 - 2\int_X \dd x\,\rho^t(x) W^t(x).
  \label{eq:func_dif}
\end{equation}
The evolutions of the Wigner function $W^t(x)$ and of the classical density $\rho^t(x)$ are both unitary so $L^2$ norm of both functions is conserved: $\|\rho^t(x)\|_2^2=\|\rho^0(x)\|_2^2$ and $\|W^t(x)\|_2^2=\|W^0(x)\|_2^2$. The only time evolving quantity in the expression (\ref{eq:func_dif}) is the overlap of the Wigner function $W^t$ and the classical density $\rho^t$ named
the {\it quantum-classical fidelity} (QCF):
\begin{equation}
  F(t) = \int_X \dd x\, \rho^t(x) W^t(x).
  \label{eq:basic_qcf}
\end{equation}
The QCF (\ref{eq:basic_qcf}) is the measure for the quality of the CQC. The $F(0)$ is the maximal value of $F(t)$ on Euclidean phase space, or is maximal in the limit $\hbar\to 0$ for compact phase spaces. Therefore, in order to study CQC we observe relative QCF defined as
\begin{equation}
  G(t) = \frac{F(t)}{F(0)} \le 1 + \exp(-|O(\hbar^{-1})|).
\label{eq:relative_qcf}
\end{equation}
On the Euclidean phase space we can skip the $\hbar$ corrections on the right hand side of (\ref{eq:relative_qcf}). The relative QCF $G(t)$ is the main object through which we study CQC in the following sections. The decrease of $G(t)$ from the value $1$ is understood as notable violation of the CQC.\par
An important motivation for the study of QCF comes from the studies of classical \cite{classLE} and quantum Loschmidt echoes \cite{quantLE} and their correspondence. One can immediately show that a good control of QCF as defined above (\ref{eq:basic_qcf}) gives in turn a good control of the correspondence between the two Loschmidt echoes. In order to define the Loschmitd echoes, or fidelities, we have to follow two slightly different time evolutions run by two slightly different Hamiltonians. Let the classical Liouville densities and quantum Wigner functions of the {\em unperturbed} and {\em perturbed} evolutions at time $t$ be denoted as $\rho_0^t(x)$, $W_0^t(x)$, and $\rho_\epsilon^t(x)$, $W_\epsilon^t(x)$, respectively, where $\epsilon$ characterizes the size of perturbation. Then the classical Loschmit echo (CLE), or classical fidelity, and the quantum Loschmidt echo (QLE), or quantum fidelity, are defined respectively as
\begin{eqnarray}
  F^{\rm CLE}_\epsilon(t) &=& \int \dd x \,\rho^t_0(x) \rho^t_\epsilon(x),\\
  F^{\rm QLE}_\epsilon(t) &=& \int \dd x\, W_0^t(x) W_\epsilon^t(x).
\end{eqnarray}
For strongly chaotic systems, within the Ehrenfest time scales, QLE has been established to exhibit an exponential decay with the rate given by a classical Lyapunov exponent \cite{Jalabert}. On the other hand, again for sufficiently short time scales, CLE also exhibits Lyapunov decay as has been derived in \cite{veble04}. These results may suggest a simple conclusion that the Lyapunov regime of QLE is a purely classical phenomenon. Yet, there is a subtle issue of the precise time-scale (perhaps a fraction of the Ehrenfest time) up to which the connection may be established. And this connection can be given through the concept of QCF as shown below for the Euclidean geometry $X=\R^{2d}$ with $d$ degrees of freedom. In order to do that we shall make use of an elementary inequality between arbitrary quadruple of Hilbert space vectors $u,v,r,s$
\begin{equation}
  |\, \|u-v\|_2 - \|r-s\|_2 \,| \le \|u-r\|_2 + \|v-s\|_2
\end{equation}
The inequality can be proven straightforwardly by means of
standard triangle inequalities. Now, putting $u:=\rho_0^t$,
$v:=\rho_\epsilon^t$, $r:=W_0^t$, $s:=W_\epsilon^t$, and assuming
that Wigner-functions and classical densities are all square
normalized to a unit Planck cell $\|u\|_2 = \|v\|_2 = \|r\|_2 =
\|s\|_2 = (2\pi\hbar)^{-d}$  due to purity of the state, we obtain
an upper bound estimate on the difference between quantum and
classical fidelity in terms the sum of QCF of unperturbed and
perturbed dynamics
\begin{equation}
 \left|\sqrt{1-F^{\rm CLE}_\epsilon(t)} - \sqrt{1-F^{\rm QLE}_\epsilon(t)}\right|
  \le
  \sqrt{1-F^{\rm QCF}_{0}(t)} + \sqrt{1-F^{\rm QCF}_{\epsilon}(t)}.
  \label{eq:fidineq}
\end{equation}
For example, if within a certain time $t$, QCF remains close to $1$ for both evolutions, then within the same time QLE closely follows CLE.\par
In this paper we discuss CQC on Euclidean and toroidal geometry, the latter being an example of a compact phase space. Toroidal geometry is particularly useful, because in this case we know plenty of simple chaotic models with simple exact quantizations and additionally it supports a smooth and a discrete Wigner function (WF) formulations. We study CQC numerically through the QCF using the Sawtooth map and the Perturbed cat map as canonical examples for the linear-discontinuous, and nonlinear-smooth maps, respectively. The smooth WF on the torus can be interpreted as a model for WF defined on Euclidean phase space on which me make some heuristic predictions on the decay of the QCF. 

Some related comparative studies of classical and quantum phase space distributions 
in the relation to Lyapunov chaos can be found in Refs.\cite{arjendu1,arjendu2}.
\par
The paper is organized as follows. In section 2 we outline a heuristic derivation of the Lyapunov decay of QCF in the echo picture. In section 3 we give more rigorous definitions of QCF for the case of compact two-dimensional toroidal phase space and two possible distinct definitions of the Wigner function, and later on present numerical calculations for two specific models, namely Sawtooth map and Perturbed cat map. In section 4 we summarize our results and formulate conjectures for future rigorous work \cite{future}.\par
\section{Euclidean phase space and heuristics}
The Euclidean (symplectic) classical phase space is perhaps physically the most relevant geometry and is here denoted by $X = Q\times P = \R^{2d}$, where $Q=\{q\in\R^d\}$ is the configuration and $P=\{p\in\R^d\}$ is the momentum space. By using some heuristic derivation in this geometry we qualitatively explain the numerically observed behaviour of QCF (\ref{eq:basic_qcf}) for the situation, when classical and quantum system are initially described by a Gaussian packet for a generic chaotic system with effectively finite available phase space. The finiteness of the phase space assures, 
due to Poincare recurrence theorem \cite{katok97}, that quantum wave-packet 
stretched due to chaotic dynamics eventually interferes with itself causing a clear violation of the CQC. The point in time when this happens we call {\it initial break of CQC} and we are mainly interested in the development of QCF after that point in time. Without loss of generality we can restrict ourselves to time independent Hamiltonian systems. We further restrict our discussion to a simply connected chaotic component of phase space (e.g. the energy surface).\par
The classical and quantum system are defined in terms of the Hamilton function $H(x)$, $x\in X$ and the Hamilton operator $\hat H:{\cal H} \to {\cal H}$ over Hilbert space $\cal H$, respectively. We shall be working in Weyl-Wigner quantization. The classical dynamics can be described solely in terms of characteristics of the flow, namely the classical trajectories. We write a symplectic map $\phi^t:X\to X$ which represents a trajectory starting in the point $x\in X$ as function $\phi^t(x)$. The quantum dynamics does not support such a treatment and we usually refer to this property as {\it non-locality of quantum mechanics}. The classical propagation of the phase-space probability density $\rho(x)$ is defined by the Liouville equation
\begin{equation}
   \dot \rho = {\cal L} \rho\>,\quad
   e^{t {\cal L} }\rho(x) = \rho(\phi^{-t}(x)) = \rho^t(x)\>,\quad
   {\cal L} = \{H,\bullet\}\>,
   \label{eq:lioville_eq}
\end{equation}
Here we use the Liouville operator ${\cal L}$, which we express using the 
bi-directional differential $\bJ$ as
\begin{equation}
   {\cal L} = H \bJ\>,\quad
   \bJ = \sum_{ij} J_{ij} \pa_i^{\rm L} \pa_j^{\rm R}\>, \quad
   {\bf J}=\left[\begin{array}{cc} 0 & \id \cr -\id & 0 \end{array}\right]\>,
   \label{eq:lioville_op}
\end{equation}
where superscripts L (left) and R (right) indicate the direction in which the differentials act and matrix $\bf J$ is the symplectic unit. The Wigner-Weyl formalism in this geometry is well known \cite{lee95}. The Wigner function $W_\psi(x)$ of the quantum state $\ket{\psi}$ is defined by
\begin{equation}
  W_\psi (q,p) = \frac{1}{(2\pi\hbar)^d}\int \dd v\,
  e^{\ii p v/\hbar} \braket{\psi}{q + v/2}\braket{q-v/2}{\psi}\>.
  \label{eq:wigner_euclid}
\end{equation}
Similarly to classical propagation of the probability density, we can define the dynamics of the Wigner function $W^t(x)$ using the Weyl-Liouville operator ${\cal L}_{\rm w}$
introduced by Moyal \cite{lee95} as
\begin{equation}
  \dot W^t =  {\cal L}_w W^t,\qquad
  W^t(x) = e^{t {\cal L}_{\rm w}} W(x),\qquad
 {\cal L}_{\rm w} = H\, \frac{1}{\alpha} \sin(\alpha \bJ) \>,
 \label{eq:weyl-lioville_eq}
\end{equation}
where we introduce a constant $\alpha=\hbar/2$. The operators ${\cal L}$ and ${\cal L}_{\rm  w}$ are generators of unitary dynamics on $L^2(X)$ and they do not commute in general $[{\cal L},{\cal L}_{\rm w}] \neq 0$, which is essential for the properties of quantum mechanics. The coherent state with a deformation $\sigma$ in position direction and centered at the phase space point $(q_0,p_0)\in X$ is written as
\begin{equation}
  \braket{q}{q_0,p_0} = \left(\frac{\sigma}{\hbar\pi} \right)^{\frac{d}{4}}
  e^{ -\frac{\sigma}{2\hbar}(q-q_0)^2 + \frac{\ii p_0 q}{\hbar}}\>.
  \label{eq:euclid_coh}
\end{equation}
The product of uncertainties in position and momenta reads $\Delta q \Delta p = \hbar/2$. The Wigner function corresponding to the coherent state (\ref{eq:euclid_coh}) is
\begin{equation}
  W_{(q_0,p_0)} (x)  = \frac{1}{(\pi\hbar)^d}
  e^{ -\frac{\sigma}{\hbar}(q-q_0)^2 -\frac{1}{\hbar\sigma}(p-p_0)^2}\>,
  \label{eq:euclid_wig0}
\end{equation}
which is completely positive with a unit phase space integral. The functions of the form (\ref{eq:euclid_wig0}) are used as initial conditions in the evolution of classical and quantum system, represented by the Wigner functions $W^t(x)$ and the classical probability densities $\rho^t(x)$:
\begin{equation}
 W^{t=0}(x) = \rho^{t=0}(x) = W_{(q_0,p_0)} (x)\> .
  \label{eq:initial_cond_wf}
\end{equation}
By considering the time evolution of the Wigner functions and the corresponding classical density function the QCF (\ref{eq:basic_qcf}) takes in this geometry the following explicit form
\begin{equation}
   F(t) = \int_X \dd x\,
      W^0(x) [e^{-t {\cal L}} e^{t {\cal L}_{\rm w}} W^0](x)
      = \int_X \dd x\, z^0(x) z^t(x)\>,
  \label{eq:euclid_fid}
\end{equation}
which we simplify by introducing a convenient observable $z^t(x) = e^{-t {\cal L}} W^t$ named {\it classically echoed Wigner function}. The dynamics of $z^t(x)$ is defined by equations
\begin{equation}
  \dot z = {\cal R}(t) z,\qquad
  z^t =  e^{-t {\cal L}} e^{t {\cal L}_{\rm w}} z^0,\qquad
  {\cal R}(t) =
  e^{-t {\cal L}} ({\cal L}_{\rm w} - {\cal L}) e^{t {\cal L}}\> ,
  \label{eq:euclid_z_evol}
\end{equation}
with the initial condition $z^{t=0}= W^{t=0}$. The generator of evolution  ${\cal R}(t)$ is just the difference between the quantum and classical generators, ${\cal L}_{\rm w}$ and ${\cal L}$, in the
{\it classical interaction picture} and is explicitly written as
\begin{equation}
  {\cal R}(t)
   = \sum_{n=1}^\infty \frac{(-1)^n \alpha^{2n}}{(2n+1)!}
        \sum_{ij=1}^{2d} J_{ij} (\pa_i^{2n+1}H)(\phi^t(x))\,
         e^{-t {\cal L}} \pa^{2n+1}_j e^{t {\cal L}}\> .
    \label{eq:euclid_R_explicit}
\end{equation}
where we assume smallness of the semi-classical parameter
$\alpha\ll 1$. In the strong chaotic case with sufficiently simple
topology of the flow the leading contributions to ${\cal R}(t)$
can be estimated by considering that pre-factors $(\pa_i^m
H)(\phi^t(x))$ behave as stochastic variables in time and that
hyperbolicity of the flow causes the derivatives in the
interaction picture to grow exponentially with the asymptotic rate given by
maximal Lyapunov exponent $\lambda_{\rm max}$ as $\|e^{-t {\cal
L}} \pa_j e^{t {\cal L}}\| \sim \exp(\lambda_{\rm max} t)$
\cite{veble04}. Note that the observation time should be long
enough so that the unstable direction clearly develops, meaning
$\lambda_{\rm max} t \gg 1$. On the other hand, it is meaningless to treat
${\cal R}(t)$ (\ref{eq:euclid_R_explicit}) for times much larger
then $t_{\rm E} = |\log \hbar|/\lambda_{\rm max}$, because for
$t\gg t_{\rm E}$ the Wigner function already spreads over the
whole space. Taking these facts into account we find that QCF
$F(t)$ can be qualitatively estimated by
\begin{equation}
  \frac{F(t)}{F(0)}\sim e^{|\log\alpha| -\lambda_{\rm max} t}\>,
  \qquad
  1 \ll \lambda_{\rm max} t \lesssim |\log\alpha|\>.
  \label{eq:euclid_fid_res}
\end{equation}
We conclude from the formula (\ref{eq:euclid_fid_res})
that the QCF decays exponentially with the rate given by the maximal Lyapunov exponent.
In classically ergodic systems the probability density converges to the invariant density denoted by $\mu(x)$:
$$
  \lim_{t\to\infty} \rho^t(x) = \mu(x)\>,\qquad e^{t{\cal L}} \mu(x) =\mu(x)\>,
$$
and therefore one expects that QCF, for times $t \gg t_{\rm E}$, converges
to the classical ergodic average $\ave{\bullet}_{\rm e}$ of the Wigner
function
$$
  \lim_{t\to\infty} F(t) =
  \int_\chi  \dd x\, W^t(x) \mu(x) = \ave{W^t}_{\rm e}\>.
$$
The ergodic average of the Wigner function can be expected to be roughly  constant in time. This means the QCF drops to a plateau. However, for systems with relatively small Lyapunov exponents, namely if $t_{\rm E}$ is large and comparable with the Heisenberg time $t_{\rm H}\sim 1/\hbar$, QCF could experience another regime of relaxation towards the plateau related to decay of classical temporal correlation functions. This is easily understood, if we take into account two things. First, the classical density $\rho^t(x)$ converges towards the invariant measure with the rate determined by the leading Ruelle resonance $\nu$ \cite{ruelle86}\footnote{In some dynamical systems the Ruelle resonances are well established term i.e. Axiom A, geodesic flow on a manifold of negative curvature, whereas in many others, even with exponential decay of correlations, the situation is not yet clear.} with the corresponding eigenfunction $\phi_\nu(x)$ as
$$
  \rho^t(x) \approx \mu(x) + \phi_\nu(x)\ave{\phi_\nu,\rho^0} e^{-\nu t}\>,
$$
and second, that at time $t\sim t_{\rm H}$ the wave-function of a classically chaotic system is almost a random function \cite{berry77} with fluctuations on the scale of $\hbar$ and similarly for the corresponding Wigner function \cite{horvat03}. Putting this two facts together, we can estimate the asymptotic decay of QCF down to the plateau by
\begin{equation}
  F(t) - \ave{W}_{\rm e} \approx
    e^{-\nu t}\ave{\phi_\nu,\rho^0}\ave{W_\psi^t,\phi_\nu}\>.
    \label{eq:euclid_ruelle_res}
\end{equation}
By assuming that the Wigner function lives in effectively finite Hilbert space $\tilde {\cal H}$ and by introducing the projector $P:L^2(X)\to\tilde{\cal H}$ from full Hilbert space to the effective finite one, we can approximate the matrix element
$$
 |\ave{W_\psi^t,\phi_\nu}| \le \| P\phi_\nu \|_2\>.
$$
\section{QCF on the torus $\T^2$}
\subsection{Quantum mechanics over the torus $\T^2$}
The mathematical description of quantum mechanics over a compact phase space has a long history. Here we restrict ourselves to recall the main definitions and properties that are need in the exploration of quantum-classical fidelity, refereing to \cite{schwinger70} and \cite{lnp618} for details and references therein.\par
The most important consequences of doing quantum mechanics on a compact phase space such as $\T^2=[0,1]^2$ are that the inverse of the Planck's constant $\hbar$ can only take positive integer values and moreover the corresponding Hilbert space of quantum states turns out to be finite dimensional with dimension exactly $N=h^{-1}$. This is an immediate consequence of periodicity of states in both configuration and momentum representation and it will have, as we will see, important consequence on the structure and properties of both coherent states and Wigner functions. Just for the purpose of fixing the notations, let us recall briefly the construction of the Hilbert space of states and the implementation of the dynamics.\par
We start by defining the quantum Hilbert space associated to the torus phase space $\T^2$. Physically we must select wave functions satisfying the same Bloch conditions:
$$
  \psi(q + L) = e^{2\pi \ii\theta_1 L}\psi(q),\quad
  \hat\psi(p+ L) = e^{-2\pi\ii\theta_2 L}\hat\psi(p)\>,
$$
where we have used the $\hbar$-Fourier transform
\begin{equation}
  \hat\psi(p)
  =
  (\hat F_\hbar \psi)(p)\coloneq \int_{-\infty}^\infty \psi(q)\,
  \e^{-\ii qp/\hbar}\,\frac{\rmd q}{\sqrt{2\pi\hbar}} \>.
\label{e:hbar-FT}
\end{equation}
For this short presentation of the quantum kinematics we set $L=1$ and use periodic boundary conditions $\theta_1=\theta_2=0$ to keep the notations simple. Later on, in order to simplify the quantum propagator associated to the sawtooth map, we will consider $L$ even and it is important to stress that all the considerations we make here can be naturally extended to the more general case of a torus with arbitrary integer sides.\par
Note that we can characterize periodic wave functions using the Heisenberg translations: for any $\hbar\in (0,1]$, we consider the quantum translations (elements of the Heisenberg group) $\hat T_{\bv} = \e^{\ii(v_2\hat q -v_1\hat p)/\hbar}$, $\bv\in\R^2$, acting on $L^2(\R)$, obtained by the canonical quantization of the linear translations in phase space, with infinitesimal generators $\hat q$ and $\hat p$. By extension they also act on a larger set of wave functions, not necessarily in $L^2(\R)$, such as periodic functions or infinite linear combinations of delta functions over a discrete periodic lattice. In this {\it distributions space} ${\cal S}'(\R)$, we define
the space of invariant wave functions
$$
  \cH_\hbar=\{\psi\in\cS'(\R):\hat T_{(1,0)}\psi =\hat T_{(0,1)}\psi=\psi\}\>.
$$
These are distributions $\psi(q)$ which are $\Z$-periodic, and such that their $\hbar$-Fourier transform (\ref{e:hbar-FT}) is also $\Z$-periodic. One easily shows that this space is nontrivial iff $(2\pi\hbar)^{-1} = N\in\N$, which is assumed from now on, and is denoted by $\hn=\cH_\hbar$. It forms a $N$-dimensional vector space of distributions admitting a ``position representation''
\begin{equation}
  \psi(q) =
  \frac{1}{\sqrt{N}}\sum_{j=0}^{N-1}\sum_{\nu\in\Z}\psi_j
  \delta\left[q-\frac{j}{N}-\nu\right],
\quad
\ket{\psi} = 
  \sum_{j=0}^{N-1}\psi_j\,\ket{q_j} \>,
  \label{e:q-basis}
\end{equation}
where each coefficient $\psi_j\in\C$. Here we have denoted by $\{\ket{q_j}\}_{j=0}^{N-1}$ the canonical (``position'') basis for $\hn$. In the same way as we produce generic quantum states over the torus by ``periodizing''  quantum states over the line, we will construct the basic objects of our study, such as quantum propagators, coherent states, Wigner functions, etc... by again ``projecting'' down to the torus the similar object on the line or on the plane. In particular let us define the ``projector'' on $L^2(\R)$
\begin{equation}
  \hat P_{\T^2}
  = \sum_{\bm\in\IZ^2} (-1)^{N m_1 m_2}\,\hat T_{\bm}
  = \Big(\sum_{m_2\in\Z}\hat T_{(0,m_2)}\Big)\;
    \Big(\sum_{m_1\in\Z}\hat T_{(m_1,0)}\Big)\>.
  \label{e:projector}
\end{equation}
Then $\hn = \hat P_{\t2}\left(\cS'(\R)\right)$ and we will use in
particular this projector for defining coherent states on $\T^2$. Given now a classical smooth observable $f\in C^{\infty}(\T^2)$, we want to describe how to associate to $f$ a suitable quantum operator $\hat{f}$ on $\hn$. On the ``usual'' planar $\R^2$ case, there are two main quantization methods that can be both adapted to the torus: the Weyl and the (positivity preserving) anti-Wick quantization based on coherent states. For the purposes of this paper, we will here briefly describe the first one, refereing to \cite{mirkobaker} for
what concern the second one.
Note first that any smooth function $f(q,p)$ on the (unit) torus can be considered as a double periodic function over the plane $\R^2$ and it can be expressed via the discrete Fourier transform:
\begin{equation}
  f(q,p)=\sum_{\bk\in\Z^2}\tilde{f}(\bk) e^{ \ii 2\pi (q k_2 - p k_1)}.
  \label{fourier}
\end{equation}
The Weyl quantization of this function is the following operator:
\begin{equation}
  \hat{f}
   =\sum_{\bk\in\Z^2}\tilde{f}(\bk)\,T(\bk),\quad\mbox{where }
  T(\bk)=\hat{T}_{\bk/N}.
\end{equation}
It is immediate to see that $\hat f \hn\subseteq \hn$. The quantization map $f\to\hat f$, due to the finite
dimension of the quantum Hilbert space, turns out to be a {\it non} invertible transformation. In fact, because of Heisenberg relations $T(\bk + \bm N)=(-1)^{k_2 m_1-m_2 k_1}\, T(\bk)$, we have the periodic
relation $ T(\bk+ 2\bm N)= T(\bk)$. Namely, letting $\Z_{2N} = \{0,1,\ldots, 2N-1\}$ and assuming $f$ smooth:
\begin{equation}\label{2Nlattice}
  \hat f
  =\sum_{\bk\in\Z_{2N}^2}\left(\sum_{\bl\in\Z^2} \tilde{f}(\bk+2\bl N)\right)\,T(\bk).
\end{equation}
%
This, together with (\ref{fourier}) and fast decay of the Fourier coefficients,  immediately implies
$\hat{f} = \hat{g}$ for any given two smooth classical observable $f,g\in C^{\infty}(\T^2)$ that agree on the $2N$-grid
classical lattice in phase space : $f(\frac{n}{2N},\frac{m}{2N}) = g(\frac{n}{2N},\frac{m}{2N})$, $\forall m,n\in\Z_{2N}$. As it is discussed for example in \cite{BDB}, this fact will also effect the properties of the  Wigner function $W_\Psi$ associated to a given quantum state\footnote{Defined in general for any given density matrix.} $\Psi\in\hn$, that nevertheless will share the fundamental phase-space distribution property:
\begin{equation}
\bra{\Psi}\hat{f}\ket{\Psi}\,=\,N\,\int_{\T^2} W_\Psi(q,p)\,f(q,p)\,\dd q
\dd p.
\label{wigner}
\end{equation}
Before coming back to some more details concerning Wigner functions
and coherent states on the torus, let us now briefly recall the
main definition and properties of the quantum dynamics over $\T^2$.\par
In particular, in this paper our ideas are tested using some very well  known toy models of classically uniformly hyperbolic, area preserving (i.e. symplectic) maps  over $\T^2$, such as  the Sawtooth maps and Perturbed cat maps. Again, here we restrict ourself to recall the main properties needed for our considerations, refereing the reader to the cited references for additional mathematical details. While the first model (Sawtooth map) deals with piecewise linear discontinuous maps that develop diffractions effects during the evolution of given initial quantum states, the Perturbed cat maps are a prototype of non linear smooth chaotic systems and we are interested to explore the effects on QCF of both of these phenomena.\par
For any given $K\in\R$, the classical Sawtooth map is defined on the torus ${\cal T} = [0,1]\times[0,L]$, $L\in\N$, with the map $\phi_{\rm S}:(q,p)\mapsto (q',p')$:
\begin{equation}
  \begin{array}{cccl}
  p' &=& p + K q  &({\rm mod}\; L)\>,\\
  q' &=& q + p'   &({\rm mod}\; 1)\>.
  \end{array}
  \label{eq:STC_map}
\end{equation}
This map (\ref{eq:STC_map}) is not smooth for $K\neq\Z$, it is ergodic and exponentially mixing with respect to the Lebesgue measure and it has a simple expression for the maximal Lyapunov exponent:
$$
  \lambda_{\rm S} = \log \left[\frac{1}{2}(K + 2  + \sqrt{K(K+4)})\right]
$$
The corresponding quantum evolution operator $\hat U_{\rm S} \in \C^{NxN}$ over the Hilbert space of size $N$ is defined as \cite{sawtooth}
\begin{equation}
  \hat U_{\rm S} =
     \exp \left(-\ii\frac{\pi L}{N}\hat m^2\right)
     \exp \left(\ii\frac{\pi K}{N L}\hat n^2\right),
  \label{eq:STQ_map}
\end{equation}
where we introduce operators $\hat m \ket{p_m} = m \ket{p_m}$ and $\hat n \ket{q_n} = n \ket{q_n}$ with $n, m\in\Z_N$. Please note that for odd $N$ periodicity of free-propagation in direction of momenta is broken unless the dimension of the phase space in that direction, $L$, is even. The classical discontinuities of the dynamics strongly affect the propagation of states and the semiclassical properties of the system. Because of this,  only few partial rigorous {\it Egorov} estimates can be given in a time regime that, as we will discuss, it is really too short for fully exploring our observed numerical
results \cite{LB,sawtooth}.\par
We now turn to the so called Perturbed cat map over $\T^2=[0,1]^2$. We start from the very well known generalized Cat map $\phi:\T^2\to\T^2$ represented, in the natural coordinates $(q,p)$, by the matrix
\begin{equation}
\phi=\left(%
\begin{array}{cc}
  a & b \\
  c & d
\end{array}%
\right)\in SL(2,\Z)\>,
\end{equation}
with $\vert{\mbox Tr}(\phi)\vert=\vert a+d\vert>2$ to ensure
hyperbolicity. As a consequence, $\phi$ has two eigenvalues of the
form $e^{\pm \lambda}$, where $\lambda>0$ is the {\it uniform}
Lyapunov exponent of the map. On the plane, the two corresponding
unstable and stable eigenspaces are generated by two vectors $v^u$
and $v^s$ of the form $v^u=(1,u)$ and $v^s=(1,s)$, with quadratic
irrational slopes.  If $\pi:\R^2\to\T^2$ denotes the usual
projection $\mod 1$ on the torus, the image $\pi\left(\R\cdot
v^u\right)$ coincide with the unstable manifold $W^u(0,0)$ passing
trough the origin. Similarly this holds also for the stable
manifold $W^s(0,0)=\pi\left(\R\cdot v^s\right)$ and this linear
structure will be immediately lost when one introduce some
nonlinearity by perturbation. Because of the irrationality of the
slope, each invariant manifold is dense on the torus and this
property remains under perturbation. This implies that longer
segments of $W^u(0,0)$ becomes arbitrarily close to other pieces
of itself and to the origin\footnote{In the linear case, and then
also in the perturbed one, elementary diophantine estimates show
that two different branches of a piece of length $\ell$ of $W^u$
are at distance $\geq \frac{C}{\ell}$, for a certain constant
$C>0$.}. This phenomena clearly will produce {\it interference}
effects in the {\it long time} quantum evolution of initially
localized states and it will be responsible for the behavior of
the observed QCF at later times. Any such linear hyperbolic map
$\phi$ represents the most elementary example of two dimensional
discrete {\it highly} chaotic dynamical system: in particular it
is ergodic with respect to the Lebesgue measure, it has a dense set
of unstable periodic orbits and it is exponentially mixing. More
precisely, for any given observable $f, g \in C^k(\T^2)$ with zero
average and given finite regularity $k\in\N$:
$$
  \left\vert\int_{\T^2} f(q,p)g(\phi^{-t}(q,p))\,\dd q \dd p\right\vert\,
  \leq\,
  C e^{-t k\lambda}\parallel f\|_{C^k}\,\|g\|_{C^k}.
$$
This exponential decay of (classical) correlations and its behavior under nonlinear perturbations will clearly play a fundamental rule in understanding and explaining the numerical observed quantum-classical fidelity.

We will now consider perturbations of this linear map by composing it with a time $\mu>0$ flow of a global Hamiltonian $H(q,p)$ on $\T^2$. Assume $H$ to be a real function of the torus of the form:
$$
  H(q,p) =\sum_{n,m\in\Z}
  a_{nm} \sin (2\pi(nq+mp)) +  b_{nm} \cos(2\pi (nq+mp) ) \>,
$$
with rapidly decaying coefficients $a_{nm}$, $b_{nm}$. Then $H$ induces a vector field $X_H = \left(\frac{\rmd H}{\rmd p},-\frac{\rmd H}{\rmd q}\right)$ on $\T^2$ and a corresponding Hamiltonian flow $\theta_t:\T^2\to\T^2$, $\theta_0(q,p)=(q,p)$. We now consider the time $t=\mu > 0$ as a perturbative parameter and define a Perturbed cat map either as $\phi_\mu\coloneq \phi \circ\theta_\mu$ or $\phi_\mu \coloneq \theta_\mu\circ \phi$.
 It is known \cite{An, AA} that there exists a $\mu_{\rm max}>0$, which depends on $\phi$ and $H$, such that for values of $\mu \in[0,\mu_{\rm max})$, the map $\phi_\mu$ is still an Anosov map of the torus and moreover it is $C^0$-conjugate with the linear map $\phi$. Namely, for each $\mu$ fixed there exists a continuous Holder function $\Psi_\mu$ on $\T^2$ such that $\Psi^{-1}_\mu\circ \phi\circ\Psi_\mu = \phi_\mu$. In particular this implies that both the rigid spatial structure and the stability exponents of the periodic orbits can change drastically, but the global topological entropy remains constant.\par
\begin{figure}
\centering
\includegraphics[width=7cm]{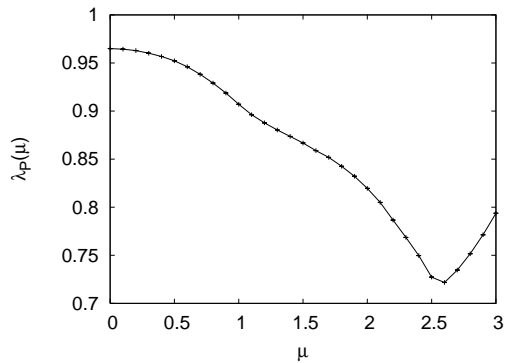}
\caption{Lyapunov exponent of the classical Perturbed cat map as a function 
of the perturbation strength, obtained as an average over 100
trajectories of length $10^6$ with random independent initial points.}
\label{pic:pcm_lyap}
\end{figure}

In this paper we consider the particular perturbations for which the Hamiltonian function $H$ of the perturbation depends only on one coordinate. In this case, the corresponding Hamiltonian flow gives rise to non-linear shears. More precisely, assume for example
$$
  H(q)= \sum_{n\in\Z} a_{n}\sin (2\pi nq) +b_{n}\cos (2\pi nq)\>.
$$
It is then easy to see that the corresponding Hamiltonian flow at
time $\mu$ is of the form:
\begin{equation}
  P_{\mu}\!  \left(\!
      \begin{array}{c}
        q \\ p
      \end{array}\! \right) \coloneq
 \left(\!
      \begin{array}{c}
        q \\ p + \mu f(q)
      \end{array}\! \right),
  \label{eq:perturb:1}
\end{equation}
a shear in momentum, where $f(q)=-\frac{\rmd H}{\rmd q}$. Similarly, a Hamiltonian of the form
$$
  H(p)= \sum_{m\in\Z} a_{m}\sin (2\pi m p) +b_{m}\cos (2\pi m p)
$$
generates a shear in position
\begin{equation}
  \label{eq:perturb:2}
  Q_{\mu}\!  \left( \!
      \begin{array}{c}
        q \\ p
      \end{array}\! \right) \coloneq
 \left(\!
      \begin{array}{c}
        q + \mu g(p) \\ p
      \end{array}\! \right),
\end{equation}
with $g(p)=\displaystyle \frac{\rmd H}{\rmd p}$. For our numerical explorations in order to simplify the numerical procedures without loosing generality, we consider the torus  ${\cal T} = [0,1]\times[0,L]$ ($L\in\N$) and we perform our calculations with the Perturbed cat map $\phi_\mu:(q,p)\mapsto (q',p')$ given by the formula
\begin{equation}
  \begin{array}{cccl}
  p' &=& q + p - \mu \sin(2\pi q)  &({\rm mod}\; L)\>,\\
  q' &=& q + p'   &({\rm mod}\; 1)\>.
  \end{array}
  \label{eq:PC_map}
\end{equation}
The hyperbolicity of this map is not uniform and the Lyapunov exponent $\lambda_{\rm P}(\mu)$ changes with perturbation strength $\mu$ as shown in the figure (\ref{pic:pcm_lyap}). Also in this case, the quantum counterpart of the classical map (\ref{eq:PC_map}) and some of its semiclassical and spectral properties have been extensively studied \cite{BDB2,MK,BDE,Nonne}. In particular the evolution operator $\hat U_{\rm P} \in \C^{N\times N}$ reads
\begin{equation}
  \hat U_{\rm P} =
  \exp \left(-\ii\frac{\pi L}{N}\hat m^2\right)
  \exp \left(\ii \frac{\pi}{N L}\hat n^2 +
             \ii \frac{N\mu}{2\pi L} \cos \left (\frac{2\pi}{N} \hat n\right)
       \right),
    \label{eq:PCQ_map}
\end{equation}

To successfully develop the problem of correspondence and to study the observed behavior of the QCF we  need to explore the theory of coherent states on $\T^2$ and their evolution under nonlinear or discontinuous hyperbolic dynamics both in configuration and phase-space representation (Wigner function). As we will briefly report here and in more mathematical details in \cite{future}, recent important estimates have been obtained in \cite{Nonne} for what concern the evolution of localized coherent states under the perturbed cat dynamics. It is worth mentioning that even if these represent the best known estimates about time propagation of states, they have a genuine semiclassical feature and they rely on a detailed control on the difference between the classical and quantum evolution (Egorov Theorem). For these reasons they do apply at a time scale where both non linear classical effects and quantum interference are negligible, allowing us to rigorously prove only the observed non decaying of the QCF for short initial time. This of course excludes the possibility of understanding the ``interesting'' part of the QCF behavior at larger time, where different mathematical approaches are needed, as we will here heuristically
discuss.
More precisely, when translated at the level of the Wigner function, these rigorous ``Egorov estimates'' allow to gain the explanation of the behavior of the QCF only up to time $T_1=T_E/2$, or $T_1=T_E/6$, depending on whether the map is linear (with possible discontinuities), or non-linear, respectively. Moreover, the breakdown of these estimates also  offer some understanding of the mechanism which produces  the appearance (at least heuristically) of the observed exponential decay controlled by the maximum Lyapunov exponent, up to the $1/N$ plateau at time $T_2 = T_1 + T_E$, as we will discuss more deeply at the end of the section.\par
Let us here briefly recall the main definitions and estimates concerning coherent states and their evolution,
refereing to \cite{Nonne} for the technical mathematical details.
As a starting point we consider a localized coherent states at the origin denoted by
$\Psi_{0,\tau}\in L^{2}(\R)$, defined as \cite{Per}
$$
  \Psi_{0,\tau}(q)
  =
  \left(\frac{\Im(\tau)}{\pi\hbar}\right)^{1/4}\,e^{-\frac{\imath\tau q^2}{2\hbar}}.
$$
Here $\tau\in\C$ is a point in the right half-plane related to the
{\it squeezing} of the coherent state. Coherent states
$\Psi_{x,\sigma}$ localized on other points $x=(q,p)$ in phase
space can then be obtained by the action of the Heisenberg
translation operators. The corresponding coherent state
$\Psi_{x,\tau,\T^2}\in\hn$ on $\T^2$ is then obtained by
projecting down the state through the operator $\hat P_{\T^2}$ as
defined in the eq.(\ref{e:projector}) \cite{mirkobaker}:
$$
  \Psi_{x,\tau,\T^2}=\hat P_{\T^2}\Psi_{x,\tau}.
$$
To be more concrete, if $\tau=\ii \sigma$ with width $\sigma>0$ then the coherent state at the point
$x=(q_0,p_0)\in\R^2$, denoted again with a small abuse of notations by $\Psi_{x,\sigma,\T^2}$,
 becomes for $\hbar=1/(2\pi N)$: %
\begin{eqnarray}
  \Psi_{x,\sigma,\T^2}\left(\frac{n}{N}\right)
  &=& \sum_{\nu\in\Z}\Psi_{x,\sigma}\left(\frac{n}{N}+\nu\right)\nonumber\\
  &=& \left(\frac{2N}{\sigma}\right)^{1/4}
      e^{\ii 2\pi n p_0}
      \sum_{\nu \in \Z}
       e^{-\frac{\pi\sigma}{N}(n - q_0 N + \nu N)^2} e^{\ii 2\pi N p_0 \nu}
  \label{coerente}
\end{eqnarray}
Note that if $q_0\in (\delta, 1 -\delta)$ then in the classical limit we can approximate the torus coherent state by the
planar one \cite{mirkobaker}:
$$
  \Psi_{x,\sigma,\T^2}\left(\frac{n}{N}\right)
  \,=\,\Psi_{x,\sigma}\left(\frac{n}{N}\right) +
  O\left((\sigma N)^{1/4} e^{-\pi N\sigma\delta^2}\right).
$$
In the study of the quantum-classical correspondence we need to compare the quantum and classical evolution of the
classical probability density corresponding to the initial phase-space representation of a given coherent state.
As we will discuss also later on, even if there is some freedom in the choice of the Wigner representation on the torus,
the previous estimate yields that the classical probability density that closely resembles the phase space representation
of the coherent state $\Psi_{x,\sigma,\T^2}$ turns out to be a periodicised Gaussian function on $\T^2$.
This last statement follows, roughly speaking, from the following
two facts \cite{mirkobaker}:
\begin{enumerate}
\item The planar coherent states, given in terms of Gaussian wavefunctions are covariant with respect the Fourier transform $\hat F_\hbar$, namely $\hat F_\hbar \Psi_{x,\sigma} = \Psi_{Fx,1/\sigma}$, where $Fx = (p_0,-q_0)$.
Now when $(2\pi\hbar)^{-1} = N$, then $\hat P_{\T^2}\hat F_\hbar = {\hat F}_{N} {\hat P}_{\T^2}$, where $\hat F_N$ again denotes the discrete Fourier transform. This shows that the above covariance still holds also after projection on the torus: $\hat F_N \Psi_{x, \sigma, \T^2} = \Psi_{Fx, \sigma^{-1}, \T^2}$.

\item The phase space representation of the coherent state, namely its (discrete) Wigner function, is essentially the Fourier transform of the overlaps between the coherent states and its translates on $\T^2$, which turns out to be related to overlap between planar coherent states, i.e. Gaussian integrals:
\begin{eqnarray*}
  \fl \bra{\Psi_{x,\sigma,\T^2}} \hat T_{\bk/N} \ket{\Psi_{x,\sigma,\T^2}}
  &=& \sum_{\bfm\in\Z^2} (-1)^{N m_1 m_2}
  \bra{\Psi_{x,\sigma}} \hat T_{\bk/N} \hat T_{\bfm} \ket{\Psi_{x,\sigma}}
  \nonumber\\
  &=& \sum_{\bfm\in\Z^2} (-1)^{Nm_1m_2+\bfm\wedge\bk}
  e^{2\pi\ii(x\wedge(\bk+\bm N))} e^{-\frac{\pi N}{2} Q_{\sigma}(\bfm+\bk/N)},
\end{eqnarray*}
\end{enumerate}
where $\bfm\wedge\bk = m_1 k_2-m_2 k_1$ and $Q_{\sigma}(\bk) = \sigma k_1^2 + \sigma^{-1} k_2^2$. As a consequences of this, it can be seen that up to exponentially small errors in $N$ (at least around the point where the coherent state is localized), we can approximate the genuine initial Wigner function of our coherent states essentially with the periodicised Gaussian function on $\T^2$
$$
  \rho_{(q_0,p_0)}(q,p) = D_N
  \left(\sum_{\nu \in \Z} e^{- 2 \pi \sigma N(q - q_0 +  \nu)^2} \right)
  \left(\sum_{\nu \in \Z} e^{- 2 \pi \sigma^{-1} N (p - p_0 +  \nu)^2} \right)
  \>.
$$
The scalar factor $D_N$ is pinned down by normalization $\int_{\T^2} \dd q\,\dd p\,\rho_{(q_0,p_0)}(q,p) = 1$, which is in the case of continuous torus $\T^2$ equal to $D_N = 2N$.\par
Let us now make a non technical analysis of the semiclassical behavior of the QCF for the short time regime $t<T_1$,
where no decay is observed. As shown in \cite{Nonne}, for short time (say $t\leq T_E/6$) the evolution of $\Psi_{x,\tau,\t2}$ is governed by the linearized dynamics around the classical trajectory of $x$. After that time, the Gaussian wavepacket starts to be seriously effected by the nonlinearity of the dynamics.
To be more precise, let us for simplicity consider an initial
 coherent state  $\Psi_{x,\tau}$
with arbitrary squeezing parameter $\tau\in\C$ and localized
around the origin (fixed point of the (un)perturbed dynamics). Let
$$
d\phi_\mu=\left(%
\begin{array}{cc}
  a_\mu & b_\mu\cr
  c_\mu & d_\mu
\end{array}%
\right)\>,
$$
be the {\it linear} hyperbolic map of the plane obtained by considering the tangent map of $\phi_\mu$ at the origin. If we consider this as a global map on $\R^2$, we denote by $V_\mu$ the corresponding quantum unitary  propagator (metaplectic representation) acting on $L^2(\R)$ is written as \cite{Fo}
$$
  \ave{q'\vert V_{\mu}\vert q} =
  \frac{1}{\sqrt{2i\pi\hbar b_\mu}}
  \exp\left(\frac{i}{2b_\mu\hbar}(d_\mu q'-2q'q+a_\mu q^2)\right)\>.
$$
Now, for "short time" $t$, the evolution of the coherent state
through the true perturbed dynamics $U_\mu^t$ can then be
approximated by an Ansatz $\Psi_t := V_\mu^t\Psi_{0,\tau} =
e^{i\theta_t} \Psi_{0,\tau_t}$, where the phase $\theta_t$ and the
"squeezing" $\tau_t$ of the new coherent states are given by the
Maslov multiplier and the homographic transformation respectively:
$$
  \theta_t = -\frac{1}{2} \mbox{arg}(a_\mu^{(t)}+\tau b_\mu^{(t)})\>,\quad
   \tau_t=\frac{c_\mu^{(t)}+\tau
  d_\mu^{(t)}}{a_\mu^{(t)}+\tau b_\mu^{(t)}}\>,
  \mbox{ where }
  \left(d\phi_\mu\right)^t =
  \left(%
  \begin{array}{cc}
    a_\mu^{(t)} & b_\mu^{(t)}\cr
    c_\mu^{(t)} & d_\mu^{(t)}
  \end{array}%
  \right)\>.
$$
Using the ``projector'' $\hat P_{\T^2}$ both  states and the propagator can be projected to the torus:
 $ V_\mu^t\Psi_{0,\tau,\T^2}\,:=\, \hat P_{\T^2} V_\mu^t \Psi_{0,\tau}$, namely $\Psi_{t,\T^2}=  \hat P_{\T^2}\Psi_t$, and the following can be proven \cite{Nonne}:
$$
\parallel U_\mu^t\Psi_{0,\tau,\T^2}\, -\,
\Psi_{t,\T^2}\parallel_{\hn}\leq C\mu \hbar^{1/2} e^{3 t\lambda_\mu},
$$
where $\lambda_\mu$ denotes the maximal Lyapunov exponent for the nonlinear Perturbed cat map.
This estimate follows from the localization  condition $\mu\Delta q(t)^3 \ll \hbar$,
where $\Delta q(t)\approx \hbar^{1/2} e^{t\lambda_\mu}$ is the width of the evolved coherent state.
Below this break time $T\approx \frac{\vert\log\hbar\vert}{6\lambda_\mu}$ the state is still very localized at the
origin, being supported in a diameter of order $\sim\hbar^{1/3}$, with no interference appearing in the phase-space
representation. In the case of piecewise linear hyperbolic maps instead, where nonlinearity is absent and only
diffraction effects at the singularities take place, techniques similar to the ones developed for the
Baker map \cite{mirkobaker} should allows a precise {\it Egorov estimates} for longer times, up to
$t\approx T_{\rm E}/2$ \footnote{For the nonlinear case at times $t>T_{\rm E}/6$ bigger than the breaking time of the linear evolution,  the states starts to be quite stretched along the unstable direction, and only suitable WKB Ansatz approximation allows one a certain control of its time evolution up to about $t\approx T_{\rm E}/2$ \cite{Nonne}}.
At this point serious interference effects between different branches of the elongate states takes place, semiclassical approximation break down and genuine quantum phenomena start to take place, as the developing of negative values in the Wigner function that are responsible for the observed time scales $T_1$, $T_2$ and the corresponding decay of the QCF as discussed in the next section.
Without entering the details here, it is important to stress that, when these estimates are translated into a phase-space representation through the Wigner function, then one can easily infer that $G(t)\approx 1$ for times below half of the Ehrenfest time for piecewise linear maps and for times below $\approx T_{\rm E}/6$ for the nonlinear smooth dynamics. This result is not surprising, at least from the physical point of view, and we would like now to develop some better understanding (at least heuristically) of the ``long time'' behavior of the QCF $F(t)$. In order to do this, we turn now to discuss the Wigner function(s) we have been using in our numerical investigations.\par
We must stress at this point that while a rigorous and unique
mathematical definition of the Weyl-Wigner formalism is now
available (see e.g. \cite{BDB}), the finite dimensionality of the
space of quantum states and the remark following
eq.(\ref{2Nlattice}) yield quite counter intuitive (at least
from a physical point of view) Wigner function defined only on a
$2N\times 2N$ discrete lattice of $\T^2$. Even if this Wigner
function does satisfies all the natural properties of a correct
phase space distribution, such as for example eq. (\ref{wigner}),
and even if it gives the expected marginals or expectation values
when integrated on linear subspaces of the phase space, it is
sometimes better to use an alternative version of the Wigner
function which offer a more intuitive interpretation of its phase
space overlap with the evolved Gaussian classical distribution
$\rho_t(q,p)$.
Even if the final numerical results do not depend on this choice, for the sake of clarity in the presentation and also for the sake of generality in the interpretation of the mechanisms yielding the exponential decay of the QCF, we discuss here the two different approaches: the discrete one (\cite{miquel02,BDB}) and the continuous one \cite{agam95}.
As we will see, we can benefit from studying both formulations of the Wigner function.
\subsection{The continuous Weyl-Wigner formalism }
\begin{figure}[!htb]
\centering
\vbox{\hskip1.4cm$t=1$}\vskip1mm
\vbox{%
\includegraphics[bb =70 65 289 288, width=4cm]{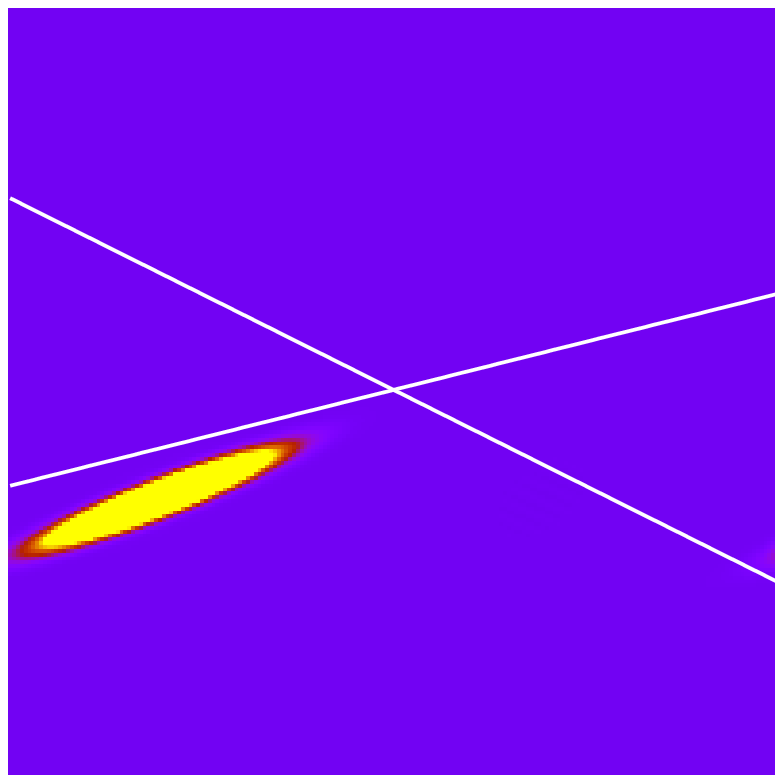}\hskip2pt%
\includegraphics[bb =70 65 289 288, width=4cm]{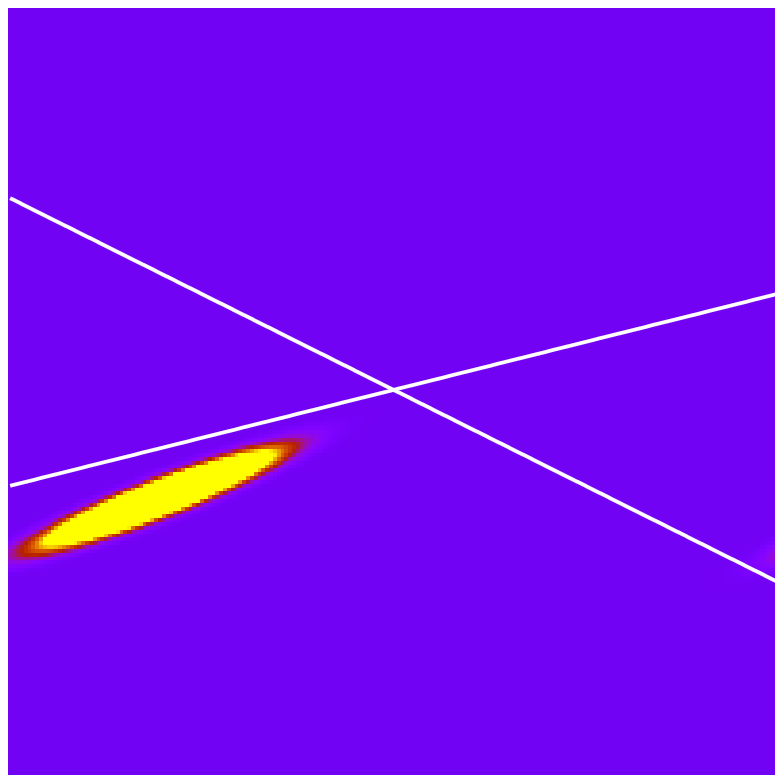}\hskip2pt%
\includegraphics[bb =70 65 289 288, width=4cm]{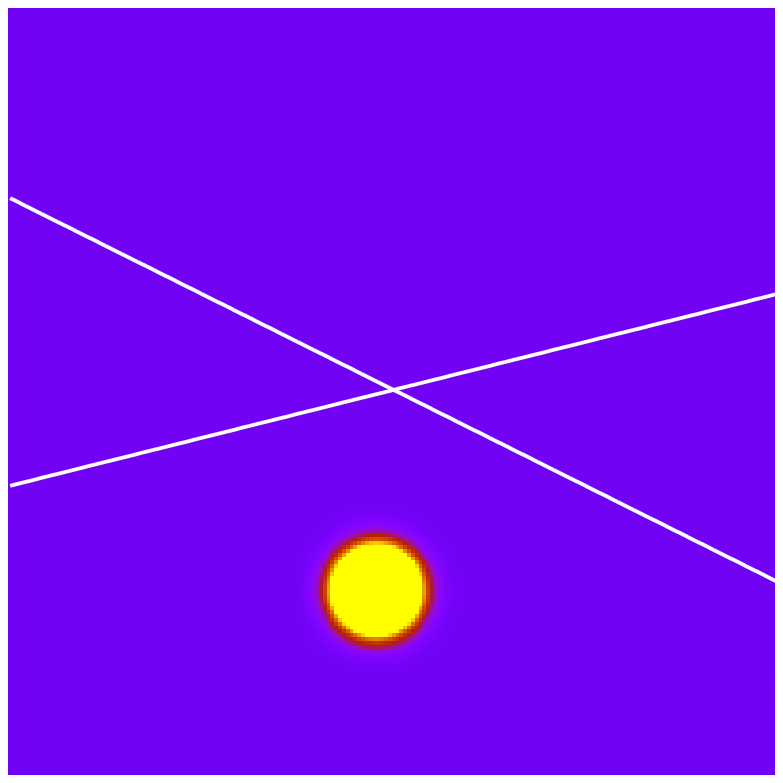}}%
\vskip-2mm\vbox{\hskip1.4cm$t=2$}\vskip1mm
\vbox{%
\includegraphics[bb =70 65 289 288, width=4cm]{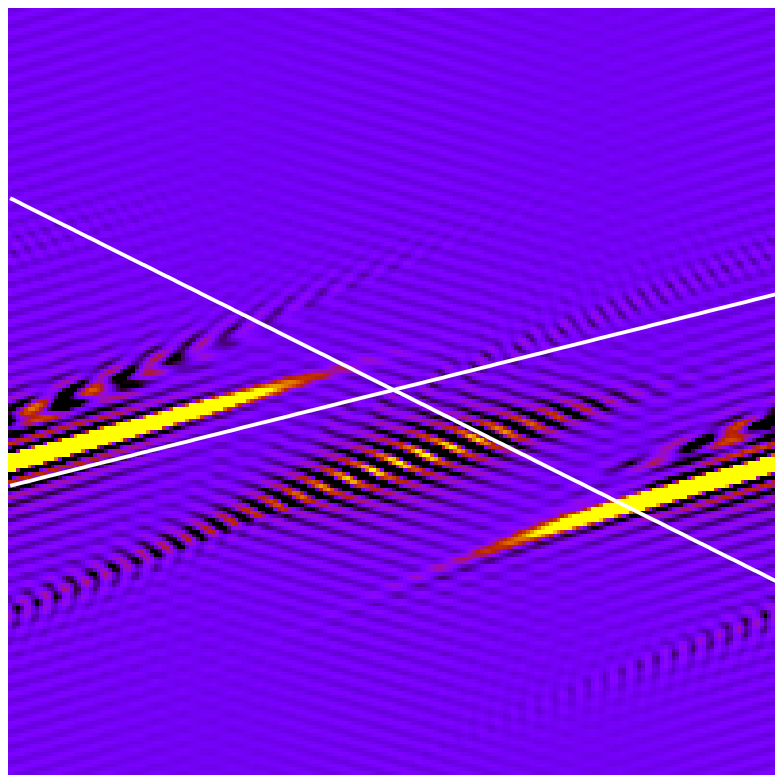}\hskip2pt%
\includegraphics[bb =70 65 289 288, width=4cm]{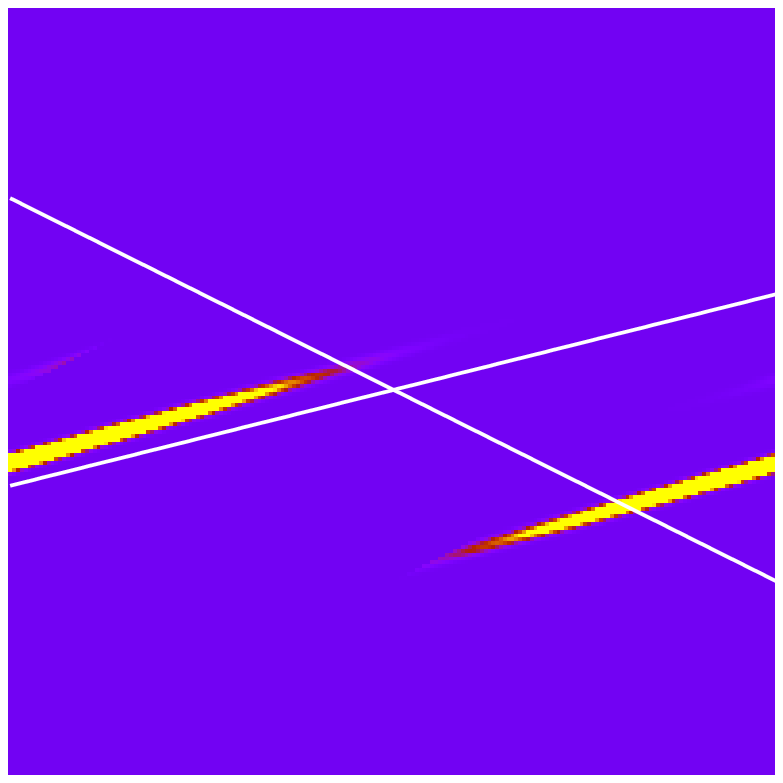}\hskip2pt%
\includegraphics[bb =70 65 289 288, width=4cm]{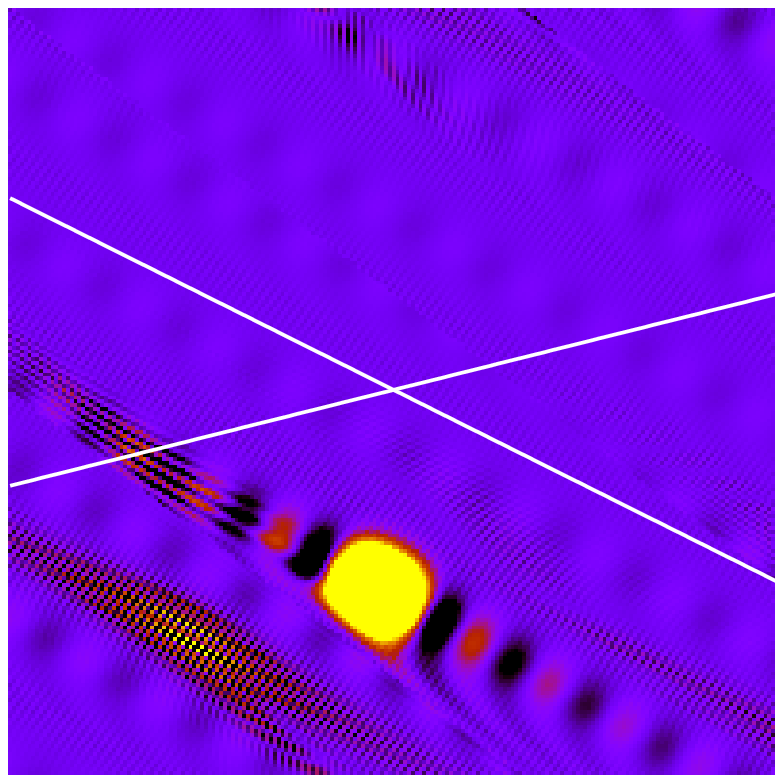}}%
\vskip-2mm\vbox{\hskip1.4cm$t=3$}\vskip1mm
\vbox{%
\includegraphics[bb =70 65 289 288, width=4cm]{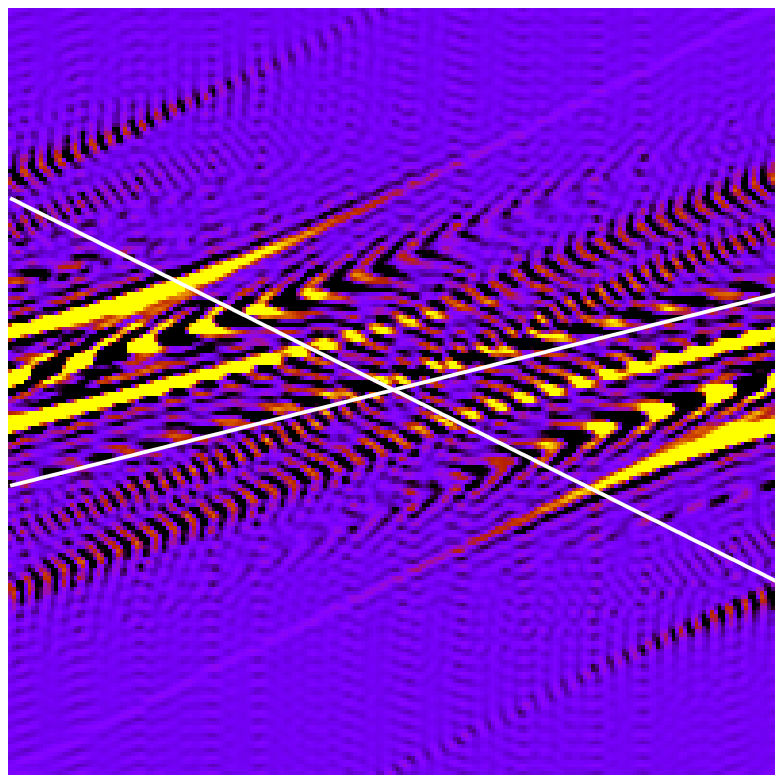}\hskip2pt%
\includegraphics[bb =70 65 289 288, width=4cm]{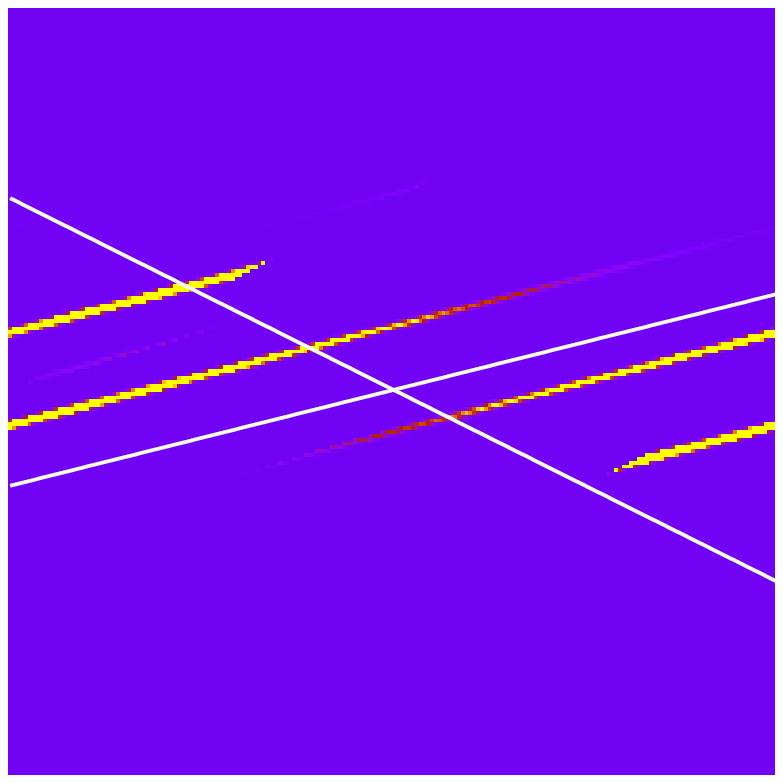}\hskip2pt%
\includegraphics[bb =70 65 289 288, width=4cm]{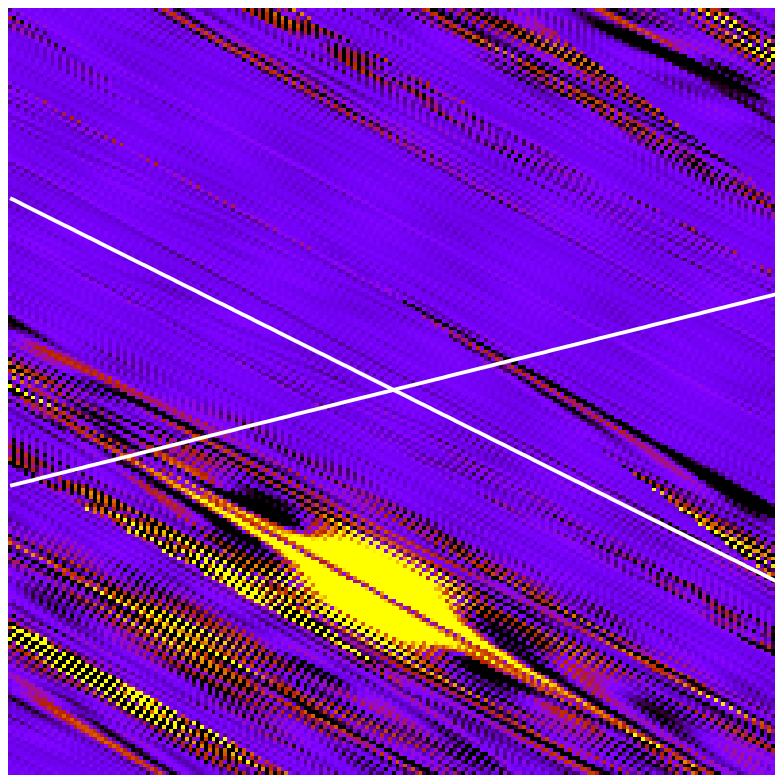}}%
\vskip-2mm\vbox{\hskip1.4cm$t=7$}\vskip1mm
\vbox{%
\includegraphics[bb =70 65 289 288, width=4cm]{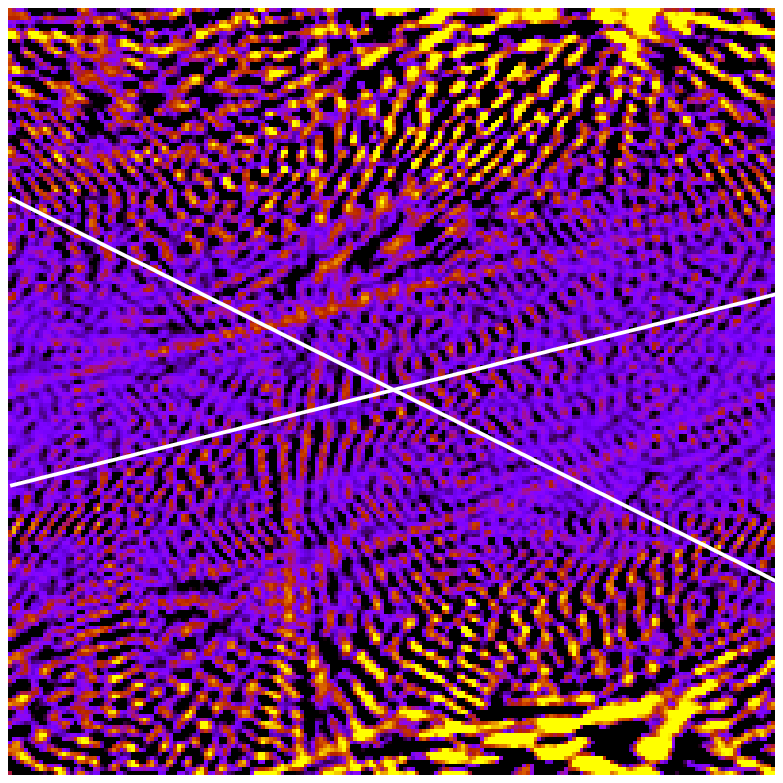}\hskip2pt%
\includegraphics[bb =70 65 289 288, width=4cm]{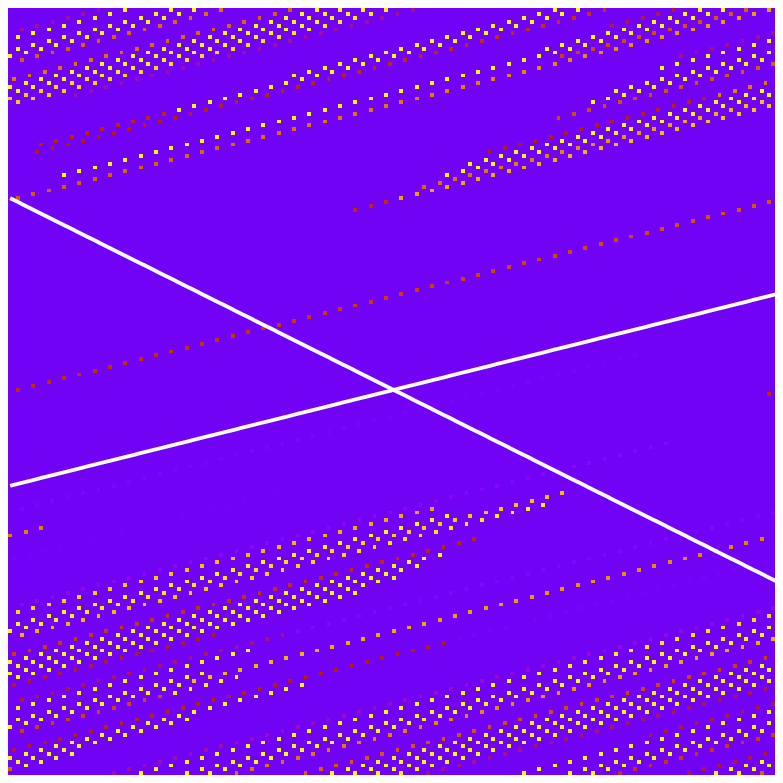}\hskip2pt%
\includegraphics[bb =70 65 289 288, width=4cm]{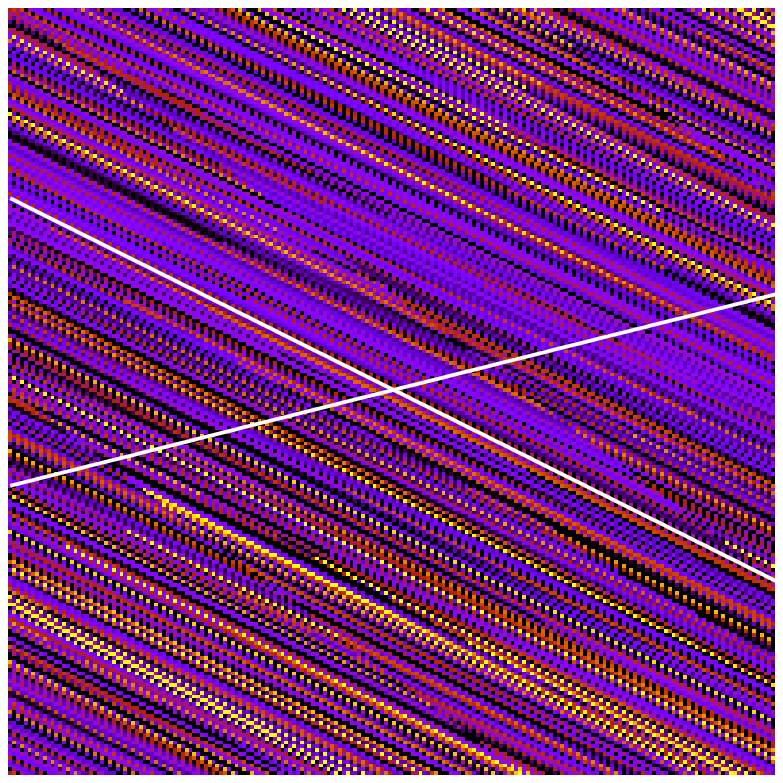}}%
\vbox{\hskip1.5cm\includegraphics[bb=70 65 382 90, width=13cm]{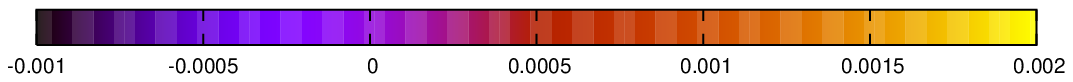}}
\caption{The density plot of the Wigner function (first column), the classical density (second column) and the classically echoed Wigner function (last column) at $t$-th step in the Sawtooth map systems at parameters $K=0.5$, $L=2$ and Hilbert space dimension $N=101$. The white lines indicate unstable and stable direction of the symplectic flow.}
\label{pic:st_dyn_wig2}
\end{figure}

The Weyl-Wigner formalism introduced by Agam\&Brenner \cite{agam95} defines the Wigner function by the following formula
\begin{eqnarray}
  W_\psi (n,m) = \frac{1}{N} \sum_{n',l= - (N-1)/2}^{(N-1)/2}
  e^{-\ii \frac{2\pi}{N} n' m}\, \tilde\delta (2l - 2n + n')\,
  \braket{q_{l+n'}}{\psi}\braket{\psi}{q_l}\>, \\
  \label{eq:WF_agam}
  (n,m)\in \Z_N^2\>, \nonumber
\end{eqnarray}
with the physical position on the torus $\T^2 = [0,1]^2$ given as $(\frac{n}{N},\frac{m}{N})\in \T^2$ and where
$$
  \tilde \delta(k)
  = \frac{1}{N} \sum_{m'=-(N-1)/2}^{(N-1)/2}
  = \frac{1}{N} \frac{\sin(\pi k /2)}{\sin(\pi k/ 2 N)}\>.
$$
This formulation of the Wigner function requests that the Hilbert space dimension $N$ is odd. This could be a deficit, as we have no concrete reason for it, except maybe that it is rather natural to assume the same number of negative as positive wave-lengths resulting in the odd number of all wave-lengths. The Wigner function given by (\ref{eq:WF_agam}) has the following properties
\begin{eqnarray*}
  \fl\hspace{5mm}
  \sum_{n=0}^{N-1} W_\psi (n,m) = |\braket{p_m}{\psi}|^2\>,\quad
  \sum_{m=0}^{N-1} W_\psi (n,m) = |\braket{q_n}{\psi}|^2\>,\quad
  \sum_{n,m=0}^{N-1} W_\psi (n,m)  = 1\>.
\end{eqnarray*}
The full Weyl-Wigner formalism that is incorporated in \cite{agam95} can be presented by using the kernel/point operator $\hat A_{nm}$ for projecting operators on the function over phase-space grid and its reconstruction from it
\begin{eqnarray*}
  \hat A = N \sum_{n,m=0}^{N-1} a_{nm} \hat A_{nm}\>,\quad
  a_{nm} = \tr\{\hat A_{nm} \hat A\}\>,\\
  \hat A_{nm} = \frac{1}{N} \sum_{n',l=-(N-1)/2}^{(N-1)/2}
  e^{-\ii \frac{2\pi}{N} n' m}\,
  \tilde\delta (2(l-n)+n')\, \ket{q_l}\bra{q_{l+n'}}\>.
\end{eqnarray*}
This type of Wigner function can be extended to the continuous
torus $(q,p)\in \T^2$ by the following definition
$$
  \tilde  W_\psi (q,p) = W_\psi \left( N p, N q\right)\>,
$$
with the basic properties
$$
  \int_{\T^2} \dd q\,\dd p\,\tilde W_\psi (q,p) = \frac{1}{N^2}\>,\quad
  \int_{\T^2} \dd q\,\dd p\,\tilde W_\psi^2 (q,p) = \frac{1}{N^3}\>,
$$
which can be treated as normalization of the Wigner function. We only use presented Wigner function in the continuous formulation as this is a more compatible object with the classical density. The classical-quantum fidelity of the evolving Wigner function $W_{\psi^t}$ and classical density $\rho^t = \rho^0(\phi^{-t}(q,p))$ on the continuous torus is defined as
\begin{eqnarray}
  F(t) &=& \int_{\T^2} \dd q\,\dd p\,\tilde W_{\psi^t} (q,p) \rho^t(q,p)\>, \\
       &=& \int_{\T^2} \dd q\,\dd p\,\tilde W_{\psi^t} (\phi^t(q,p)) \rho^0(q,p)\>,
  \label{eq:agam_fid}
\end{eqnarray}
where we use the phase space conservation of the map $\phi^t$.
Here we obtain an interesting object to discuss $W_{\psi^t}
(\phi^t(q,p))$ that we call the classically echoed Wigner function
and corresponds to $z^t(x)$ discussed in the Euclidean geometry.
By using this object we can directly observe decay of
classical-quantum correspondence on the phase space as it spreads
from original Gaussian form over the whole phase space. This is
nicely depicted in figure \ref{pic:st_dyn_wig2}, where we plot
${\tilde W}_{\psi^t} (q,p)$, $\rho^t(q,p)$ and ${\tilde
W}_{\psi^t} (\phi^t(q,p))$ parallel to each other at each time
step in the saw-tooth systems. On figures representing classically
echoed Wigner function ${\tilde W}_{\psi^t} (\phi^t(q,p))$ we can
clearly see how is this function stretched in stable directions of
the classical map as predicted by the analysis on Euclidean phase
space.\par
\begin{figure}[!htb]
\centering
\includegraphics[width=7cm]{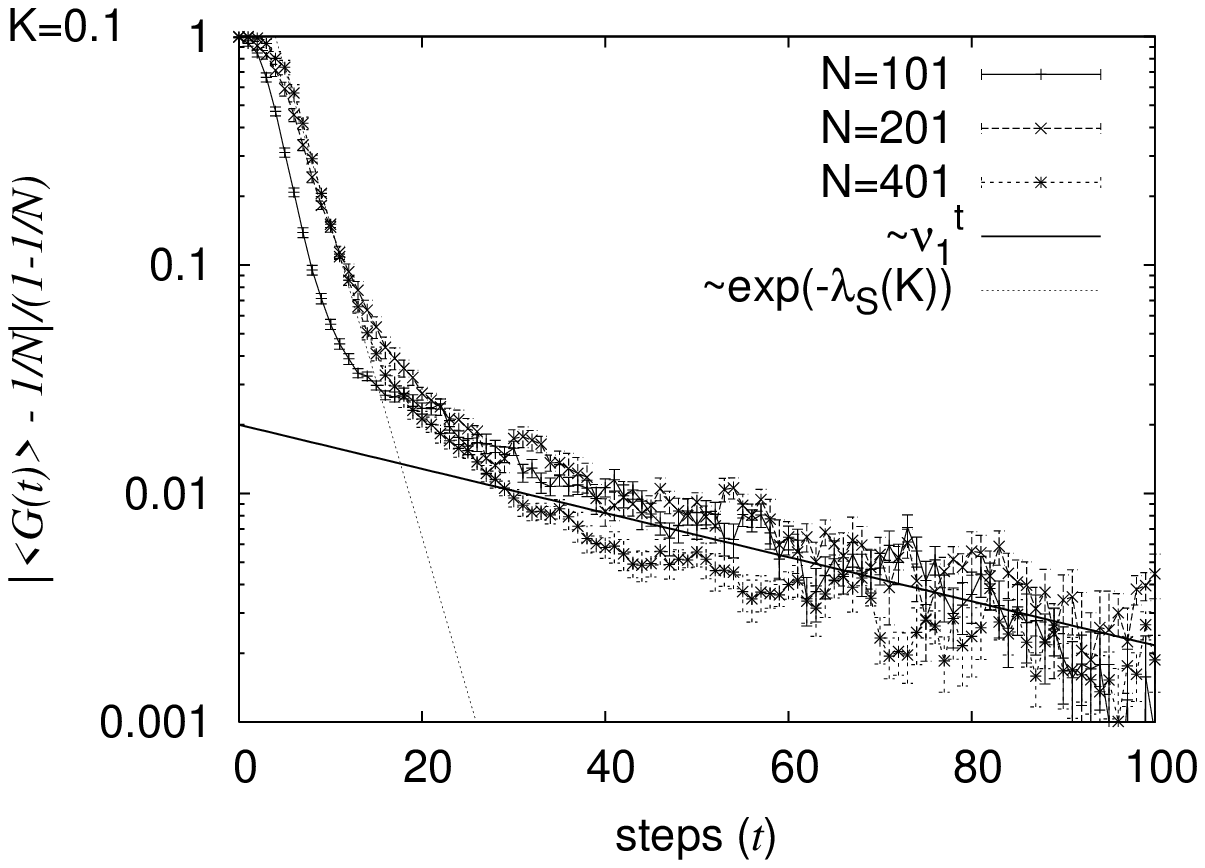}
\includegraphics[width=7cm]{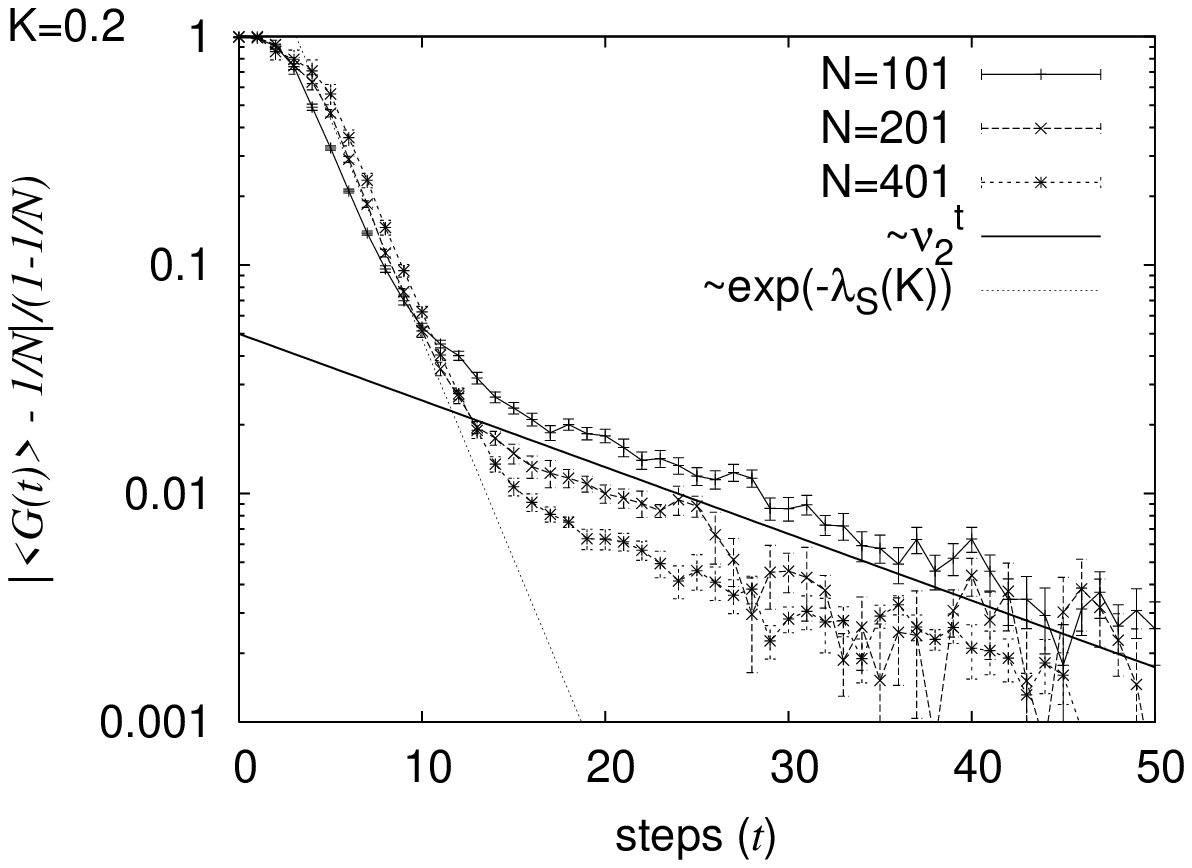}
\includegraphics[width=7cm]{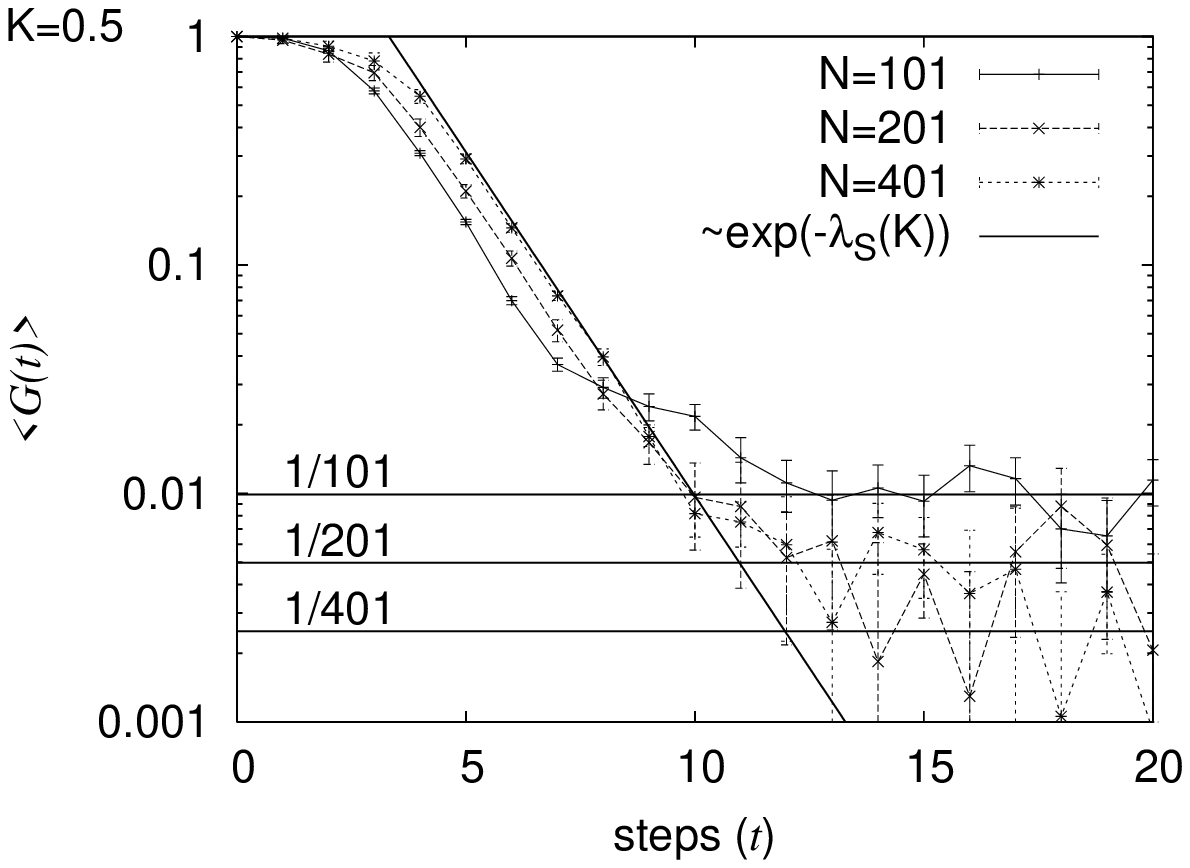}
\includegraphics[width=7cm]{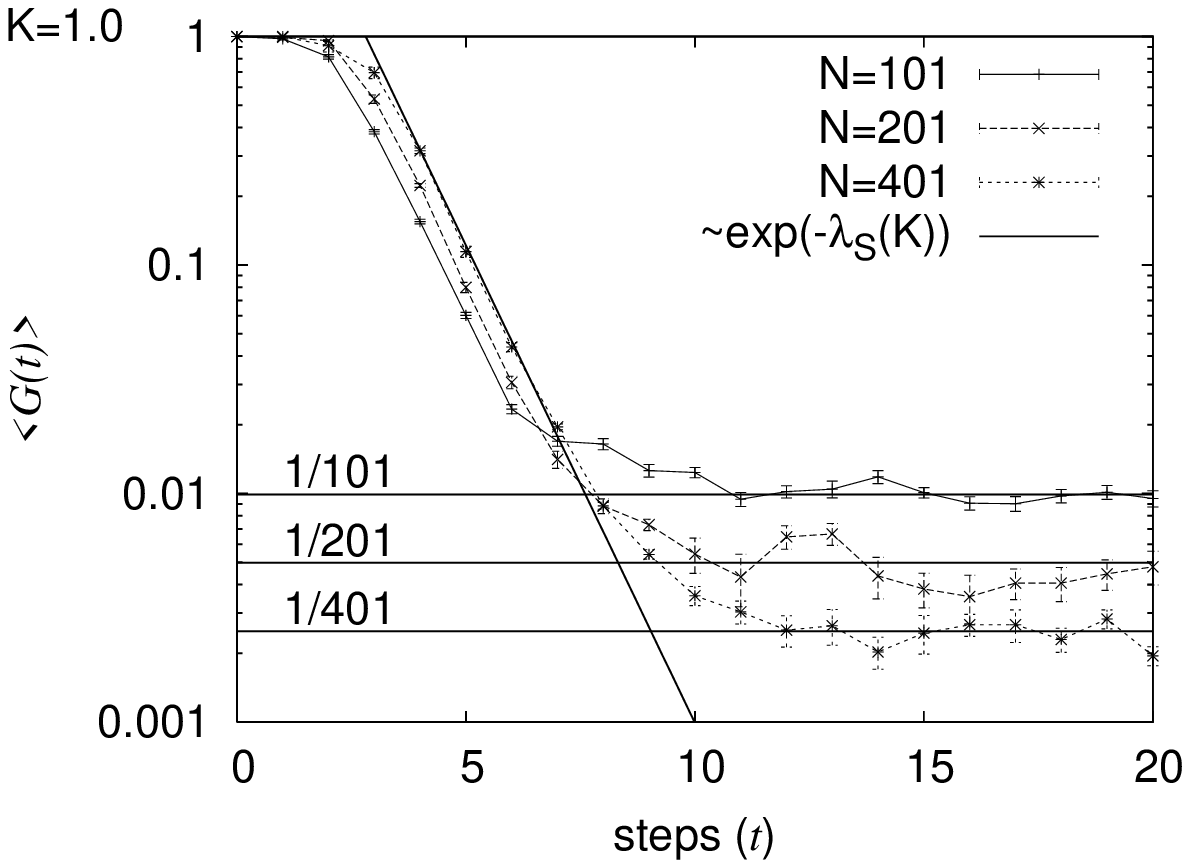}

\caption{The average QCF for the continuous Wigner function formulation in the sawtooth map at $N$ = 101, 201, 401, $L = 2$ and different $K$ = 0.1, 0.2, 0.5, 1. The average is calculated over 10 different initial Gaussian coherent states in the case $N=201,401$ and over 50 in the case $N=101$, respectively. The fitted parameters are $\nu_1$ = 0.978 and $\nu_2$ = 0.935.}

\label{pic:st_cont_wig1_1}
\end{figure}

\begin{figure}[!htb]
\centering
\includegraphics[width=7cm]{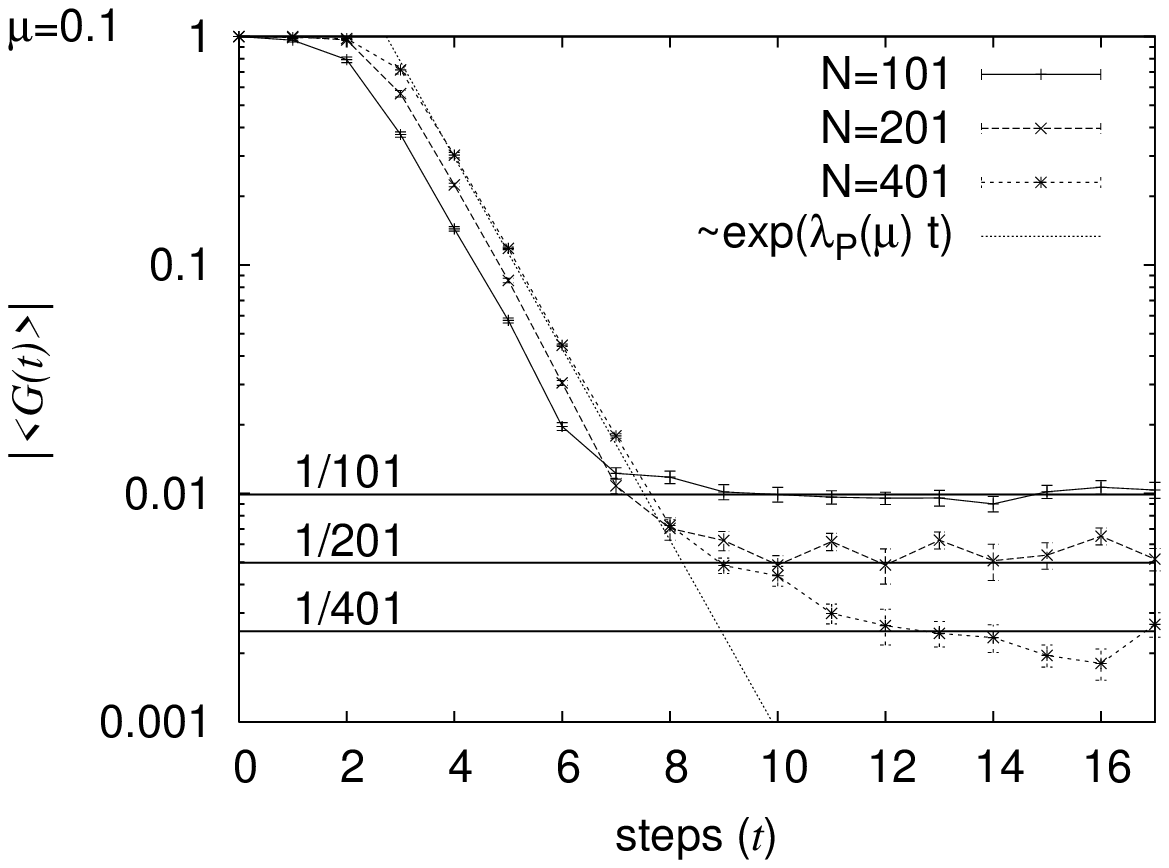}
\includegraphics[width=7cm]{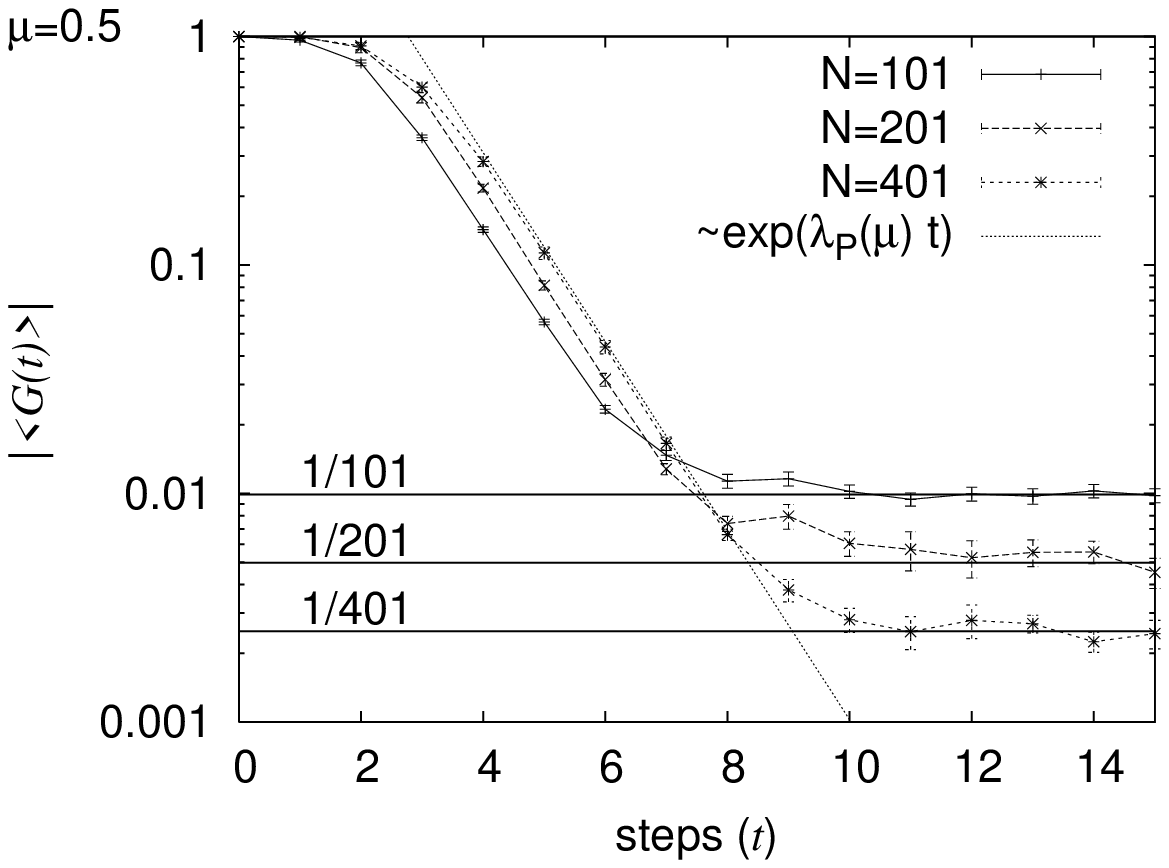}

\caption{
The average QCF for the continuous Wigner function formulation in the Perturbed cat map for $N$=101, 201, 401, $L = 2$ and different $\mu$ = 0.1, 0.5  (left,right). The Lyapunov exponents here in use are $\lambda_{\rm P}(0.1)\doteq 0.964$ and  $\lambda_{\rm P}(0.5)\doteq 0.952$. The averaging is done as for results in the figure \ref{pic:st_cont_wig1_1}.}
\label{pic:pcm_cont_wig1}
\end{figure}

We numerically calculate the quantum-classical fidelity $F(t)$ or its relative value $G(t)=F(t)/F(0)$ for the Sawtooth map system and Perturbed cat map. The results for different parameters are presented in the figures \ref{pic:st_cont_wig1_1} for the Sawtooth map and in the figure \ref{pic:pcm_cont_wig1} for perturbed cat map. The results are averaged over different positions of initial Gaussian distributions to get rid of features depending on individual initial states. In all results we can clearly see the exponential decay of relative QCF $\ave{G(t)} \sim e^{-\lambda_{\rm max}t}$ with the rate equal to the maximal Lyapunov exponent $\lambda_{\rm max}$ down to the ergodic plateau $\ave{G(t)}_{\rm e} = \frac{1}{N}$. An exception is the case of the Sawtooth system at small parameters $K$, where the expected break time $t_{\rm E}$ is rather big and we can also experience the correlation decay. Obtained result are all in the frame of presented theory for the Euclidean phase space. Note that in order to obtain reliable numerical results we have to use more then $N^2$ points in integration over the phase space, where $N$ is the Hilbert space dimension.
\subsection{Discrete Weyl-Wigner formalism }
\begin{figure}[!htb]
\centering
\includegraphics[width=7cm]{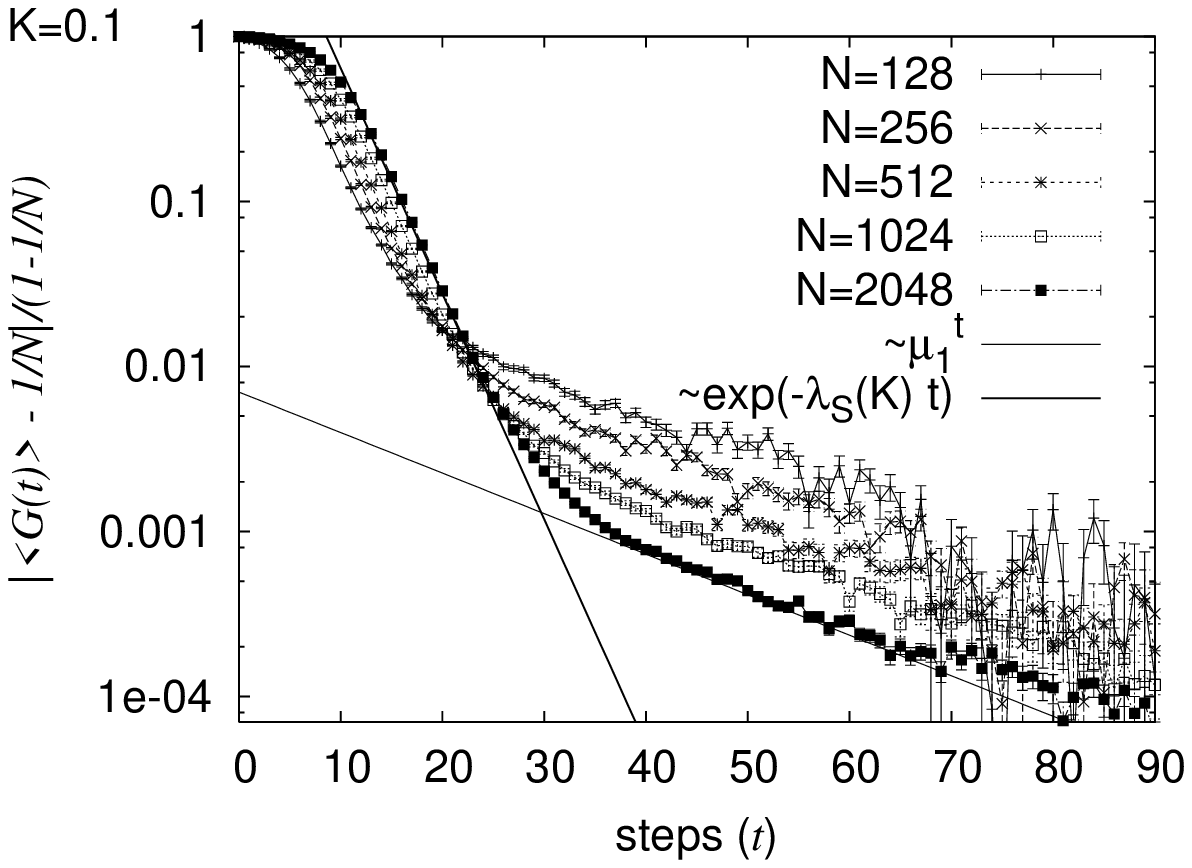}
\includegraphics[width=7cm]{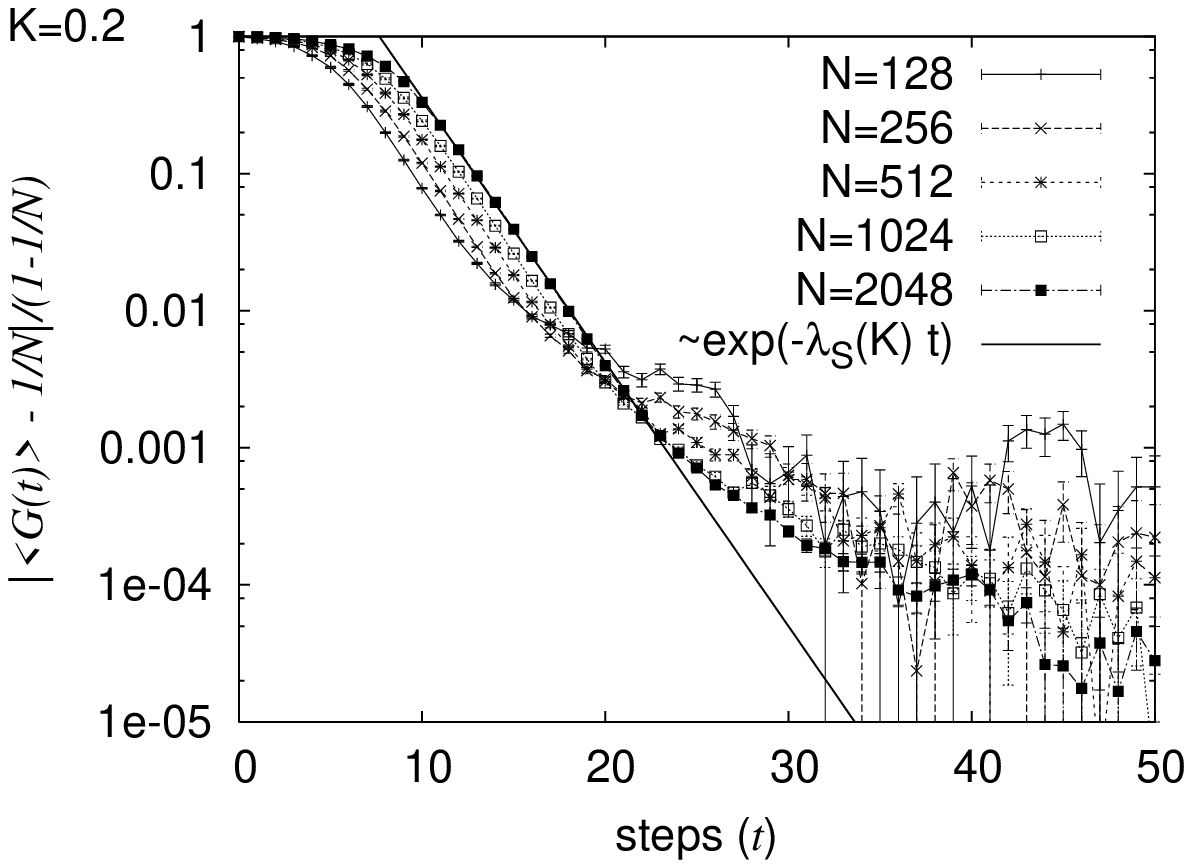}
\includegraphics[width=7cm]{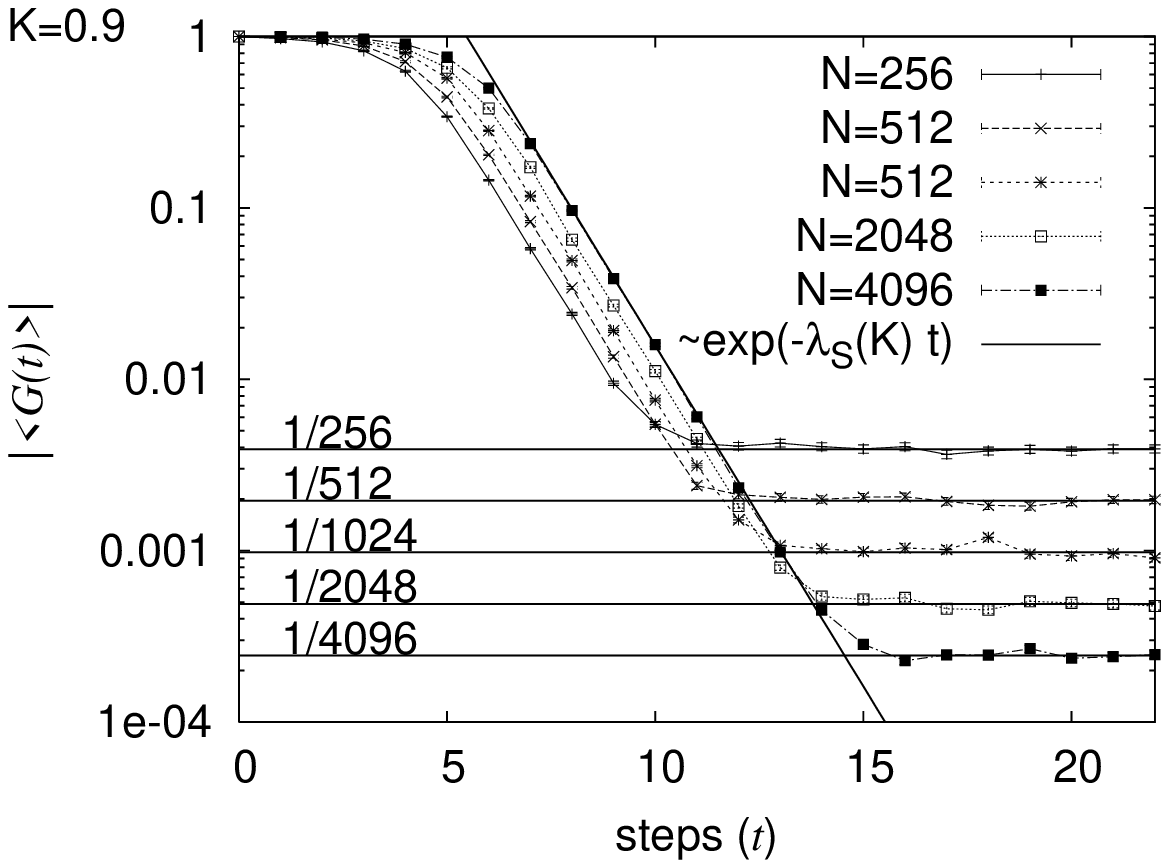}
\includegraphics[width=7cm]{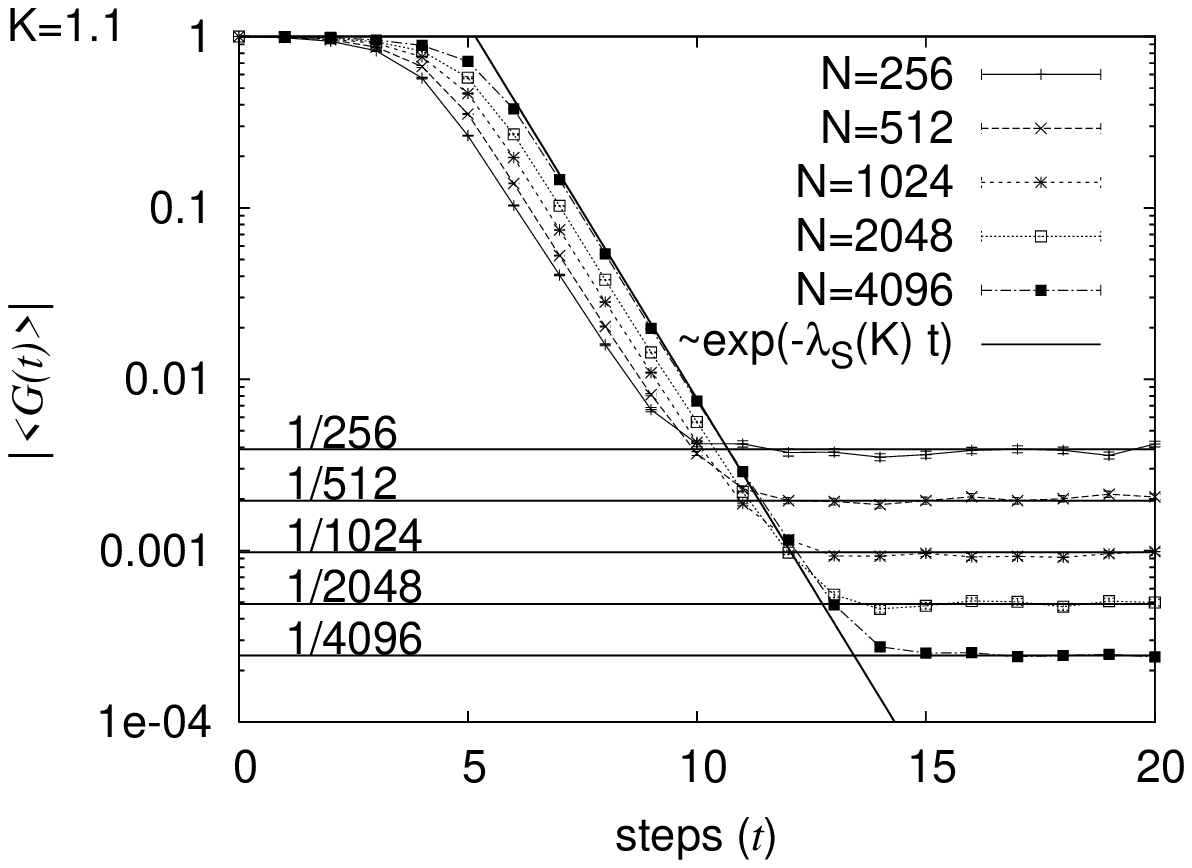}
\caption{Time evolution of the average QCF for the discrete Wigner function formulation on the lattice $2N\times2N$ in the Sawtooth map at $L=1$ for $K$=0.1, 0.2, 0.9, 1.1. The fitted paramater $\mu_1$ = 0.945. The average is taken over 500 initial Gaussian coherent states.}
\label{pic:st_discete_wig2}
\end{figure}

\begin{figure}
\centering
\includegraphics[width=7cm]{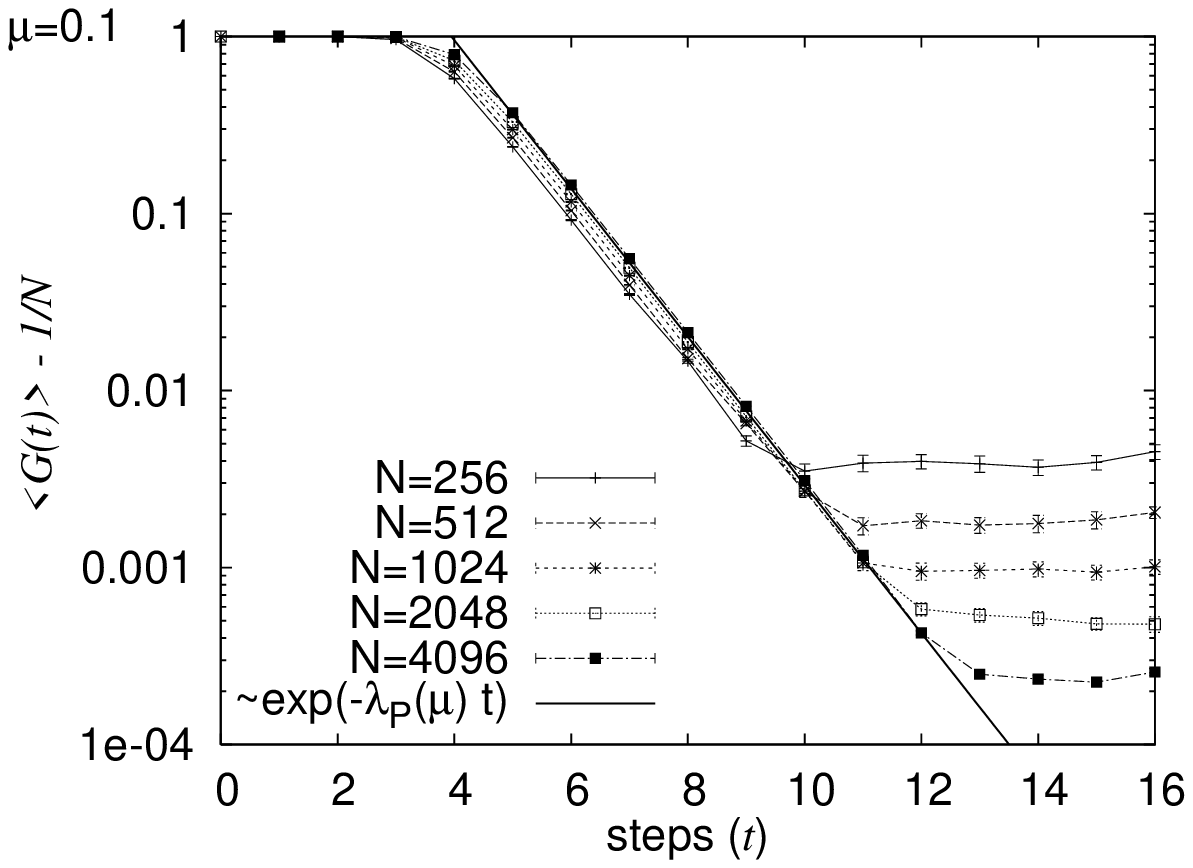}
\includegraphics[width=7cm]{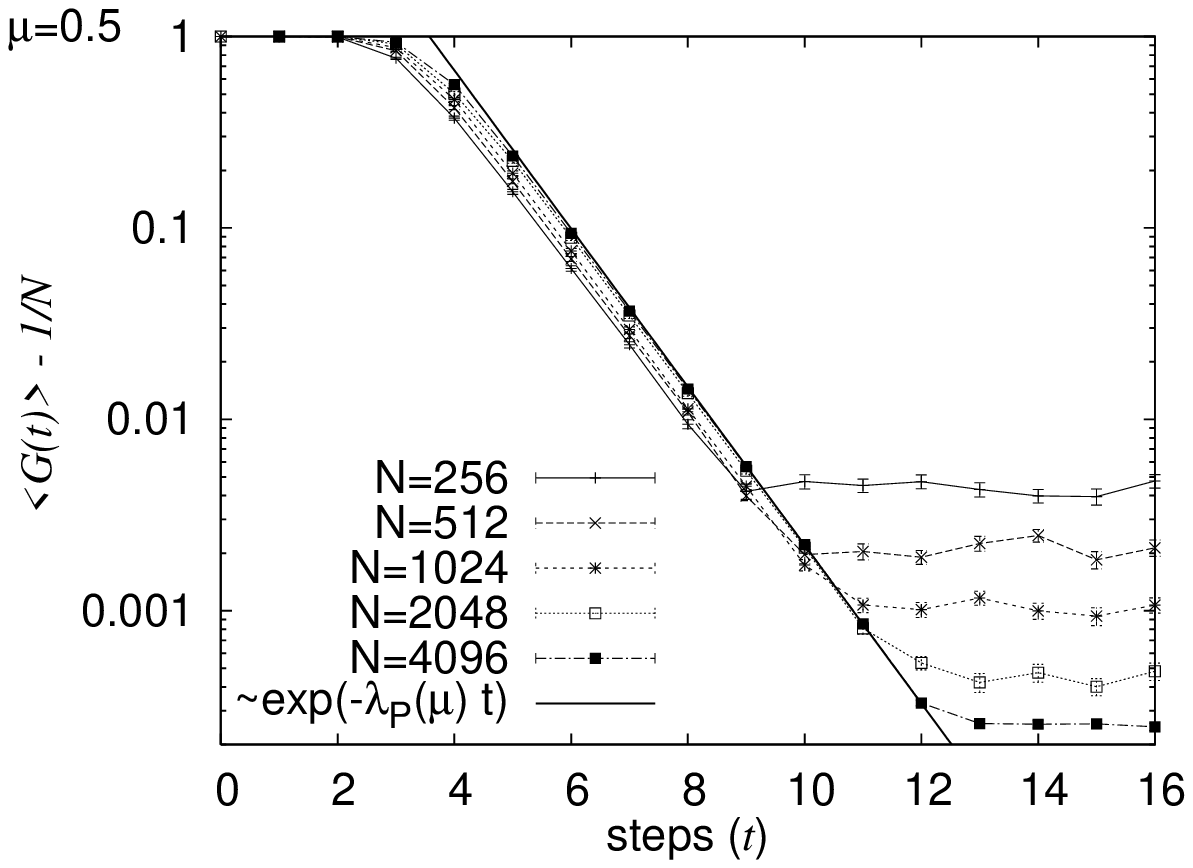}
\caption{The average QCF for the discrete Wigner function formulation in the Perturbed cat map at $L = 1$ and different values of the parameter $\mu$ = 0.1, 0.5 (left, right). Lyapunov exponents are the as used in figure \ref{pic:pcm_cont_wig1}. The average is calculated over 100 different initial Gaussian coherent states.}
\label{pic:pcm_discrete_wig2}
\end{figure}

The other Weyl-Wigner formalism that we use is described by Miquel et.al.\cite{miquel02}, by which the Wigner function reads
\begin{equation}
  W_\psi(n,m) = \frac{e^{\frac{\ii\pi}{N}nm}}{2N} \sum_{k=0}^{N-1}
  \braket{\psi}{q_{n-k}}\braket{q_{k}}{\psi} e^{-\ii \frac{2\pi}{N} k m},
  \quad (n,m) \in \Z_{2N}^2,
  \label{eq:WF_miquel}
\end{equation}
where the physical position on the torus $\T = [0,1]^2$ is given with $(\frac{n}{2N},\frac{m}{2N})\in \T^2$. Even if being defined on a $2N\times 2N$ lattice, $W_\Psi$ turns out to be completely determined by $N^2$ of its values, as the following symmetry shows:
$$
  W_\psi(n + \sigma_q N,m + \sigma_p N) =
  W_\psi(n,m) (-1)^{\sigma_q\, m + \sigma_p\, n + \sigma_q \sigma_p N}
$$
with $\sigma_{q,p} \in \{0,1\}$. Moreover, this Wigner function satisfies all the main natural requirements for a correct phase-space representation. In particular it does have the right ``Lagrangian averages'' on the points in phase space on the lattice of size $N$:
\begin{eqnarray*}
  \fl \sum_{n=0}^{2N-1} W_\psi (n,m)
  =
  \delta_m^{2k}\cdot
  |\braket{p_k}{\psi}|^2,\quad \sum_{m=0}^{2N-1} W_\psi (n,m)
  = \delta_n^{2k}\cdot |\braket{q_kn}{\psi}|^2 ,\quad
    \sum_{n,m=0}^{2N-1} W_\psi (n,m)  = 1\>.
\end{eqnarray*}
The Weyl-Wigner formalism that produces this Wigner function is generated by a point operator with a simple algebraic form
\begin{eqnarray*}
  a_{n,m} = \tr\{\hat A  \hat A_{n,m}\}\>,\quad
  \hat A = N \sum_{n,m=0}^{2N-1} a_{n,m}\hat A_{n,m}\>,  \\
  \hat A_{n,m} = \frac{e^{\ii \frac{\pi}{N} n m}}{2N}
  \sum_{k=0}^{N-1} e^{-\ii\frac{2\pi}{N} k m } \ket{q_{n-k}} \bra{q_k}\>.
\end{eqnarray*}
In the presented formulation we perform numerical experiments considering the discrete version of the quantum-classical fidelity,
\begin{equation}
  F(t) = \sum_{n,m=0}^{2N-1}
         W_\psi^t(n,m)
     \rho^t\left(\frac{n}{2N}, \frac{m}{2N}\right)\>,\quad
  \label{eq:miquel_fid}
\end{equation}
where we sum the overlap of the Wigner function and classical density over the lattice $2N\times 2N$. This formula for fidelity (\ref{eq:miquel_fid}) has the time complexity of $O(N^3 \log (N))$ and is by a factor $O(N)$ smaller than the one using the Wigner function defined by Agam \& Brenner. This enables testing of classical-quantum correspondence and its break-down at higher dimensions. We have studied such QCF in different systems: perturbed cat-map, saw-tooth map and also baker map and general linear skew shift (not reported here)
\begin{eqnarray*}
  \fl \hspace{5mm}
  \sum_{n=0}^{N-1} W_\psi (n,m) = |\braket{p_m}{\psi}|^2\>,\quad
  \sum_{m=0}^{N-1} W_\psi (n,m) = |\braket{q_n}{\psi}|^2\>,\quad
  \sum_{n,m=0}^{N-1} W_\psi (n,m)  = 1\>.
\end{eqnarray*}
The results are shown in the figure \ref{pic:st_discete_wig2} for the Sawtooth map and in the figure \ref{pic:pcm_discrete_wig2} for Perturbed cat map. Generally, in the generic case where the lattice is not conserved under the classical map, and the trajectories starting at lattice points are scattered over the whole phase space, we get the results for fidelity which look asymptotically (as $N\to\infty$) the same as in case of the Agam \& Brenner's definition of Wigner function. There exist also non-generic interesting cases. For instance in the Baker's map at dimension $N=2^n$, where the classical map in time flattens the lattice onto the ordinate axis. The other example is the saw-tooth map at $K\in\N$ called the cat map, which conserves the given lattice and the fidelity is constant, meaning it does not decay. This is a consequence of the fact that the cat map is Egorov exact \cite{mirkoX1}.

\begin{figure}[!htb]
\centering
\includegraphics[width=9cm]{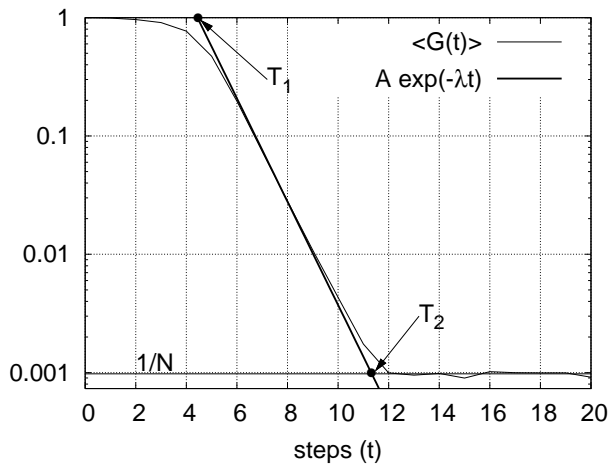}
\caption{Scheme for precise definition of two time scales, $T_1$ and $T_2$,
of decay of relative QCF $G(t)$. The thick straight line is the best 
exponential fit in the region of decay.}
\label{pic:schema_def_times}
\end{figure}

The discrete definition of the Wigner function enables to study in
more detail characteristic times of the QCF time dependence. We can
clearly recognize  two distinct time scales. The first one, denoted by $T_1$,
is the time at which $F(t)$ starts to strongly deviate from initial value
$F(0)$, entering the regime of exponential decay. The second time scale,
denoted by $T_2$, is the time when $F(t)$ exits the exponential decay and starts to relax towards the plateau. Actually, we define the times $T_1$ and $T_2$ using the average relative QCF, averaged  over different initial states $\ave{G(t)}$ and study the exponential slope of $\ave{G(t)}$: a very good agreement with the exponential fit $G_{\rm fit}(t) = \exp(T_1 -\lambda_{\rm max} t)$ is then observed for large $N$, where $\lambda_{\rm max}$ is the maximal Lyapunov exponent of the classical map. We consequently define the times $T_1$ and $T_2$ as the times when the fitting function $G_{\rm fit}(t)$ hits the initial value $G(0) = 1$ and the plateau $1/N$ respectively  as shown in the figure \ref{pic:schema_def_times}, where $N$ is the Hilbert space dimension.\par
\begin{figure}[!htb]
\centering
\includegraphics[width=7cm]{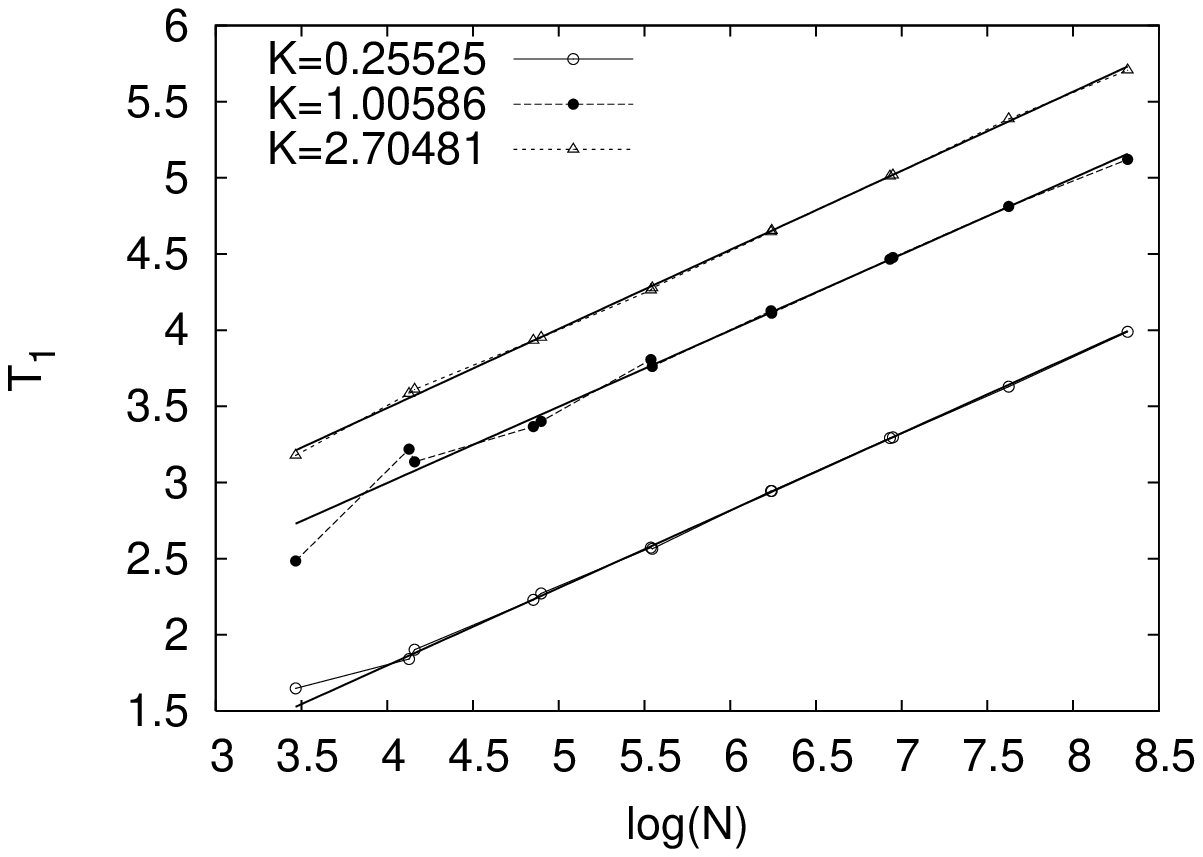}
\includegraphics[width=7cm]{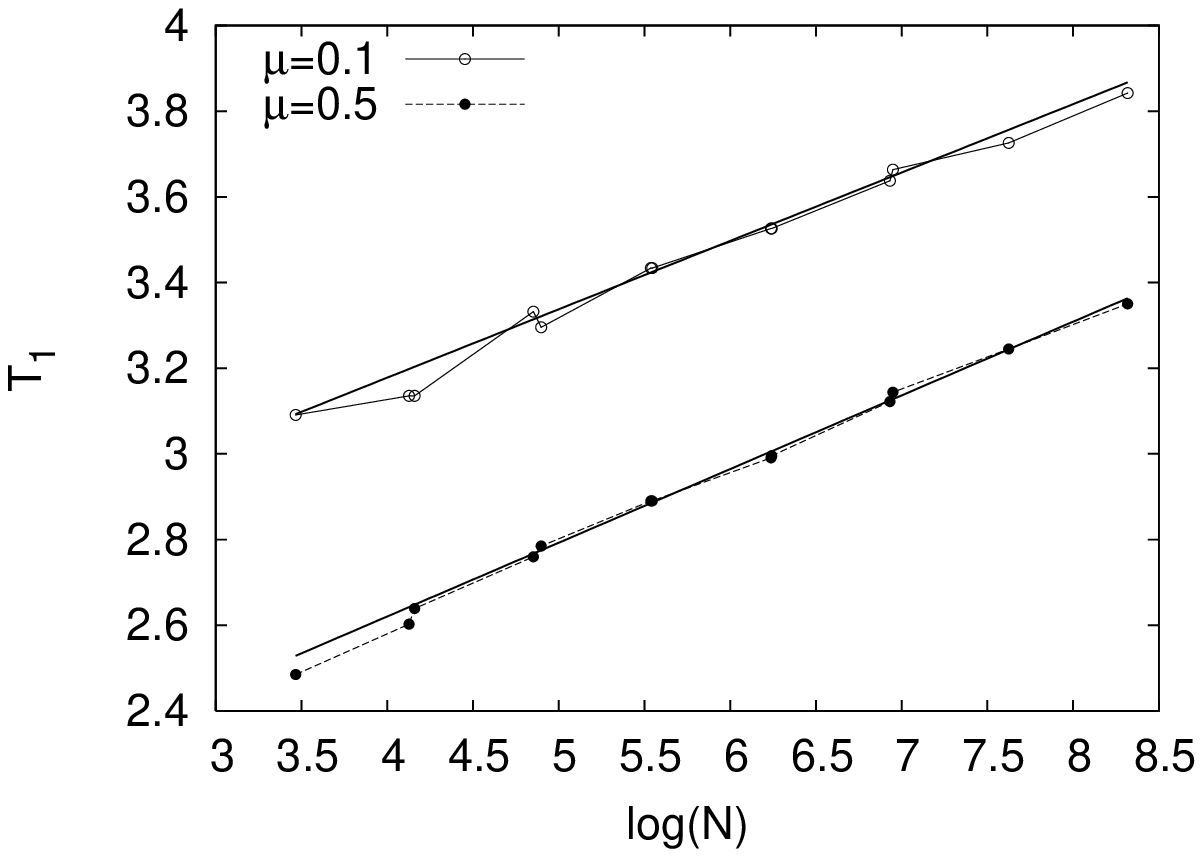}\\
\hbox to 16cm{\hfil(a)\hfil(b)\hfil}
\caption{Dependence of the time $T_1$ in decay of average relative QCF $\ave{G(t)}$ on the Hilbert space dimension $N$ for the Sawooth map (a) and Perturbed cat map at parameter values which are indicated in the figure. Inserted solid lines are linear fits: -0.234 + 0.508 $\log N$, 0.995 + 0.500 $\log N$, 1.411 + 0.519 $\log N$ in the case (a) and  2.539 + 0.159 $\log N$,  1.944 + 1.169 $\log N$ in the case (b). The average in $\ave{G(t)}$ is taken over 500 initial Gaussian packets.}
\label{pic:T1_fits}
\end{figure}

We see that the times $T_2$ and $T_1$ are connected by the relation
\begin{equation}
\lambda_{\rm max}(T_2-T_1) = \log N
\label{eq:times_scales_connection}
\end{equation}
Because of that trivial relation we discuss only the time scale $T_1$ in the following. In the simulations done with the Sawtooth map and Perturbed cat map, see figure \ref{pic:T1_fits}, we found that $T_1$ is approximately described by
\begin{equation}
 \lambda_{\rm max} T_1 \approx A \log N + B
\label{eq:times_scales}
\end{equation}
where $A$ and $B$ are constants, depending on the model but not on $N$. We are are interested only in leading term and the coefficient $A$. By numerical analysis of the models at different values of parameters we found that $A$ in Sawtooth map $T_2$ is almost independent of $\lambda_{\rm max}$ and is near to $A = 0.5 \pm 0.05 $, but in the Perturbed cat map the coefficient $A$ slightly varies with the coefficient $\mu$ and is approximately equal to $A = 0.2\pm 0.05$.\par
\begin{figure}[!htb]
\centering
\includegraphics[width=7cm]{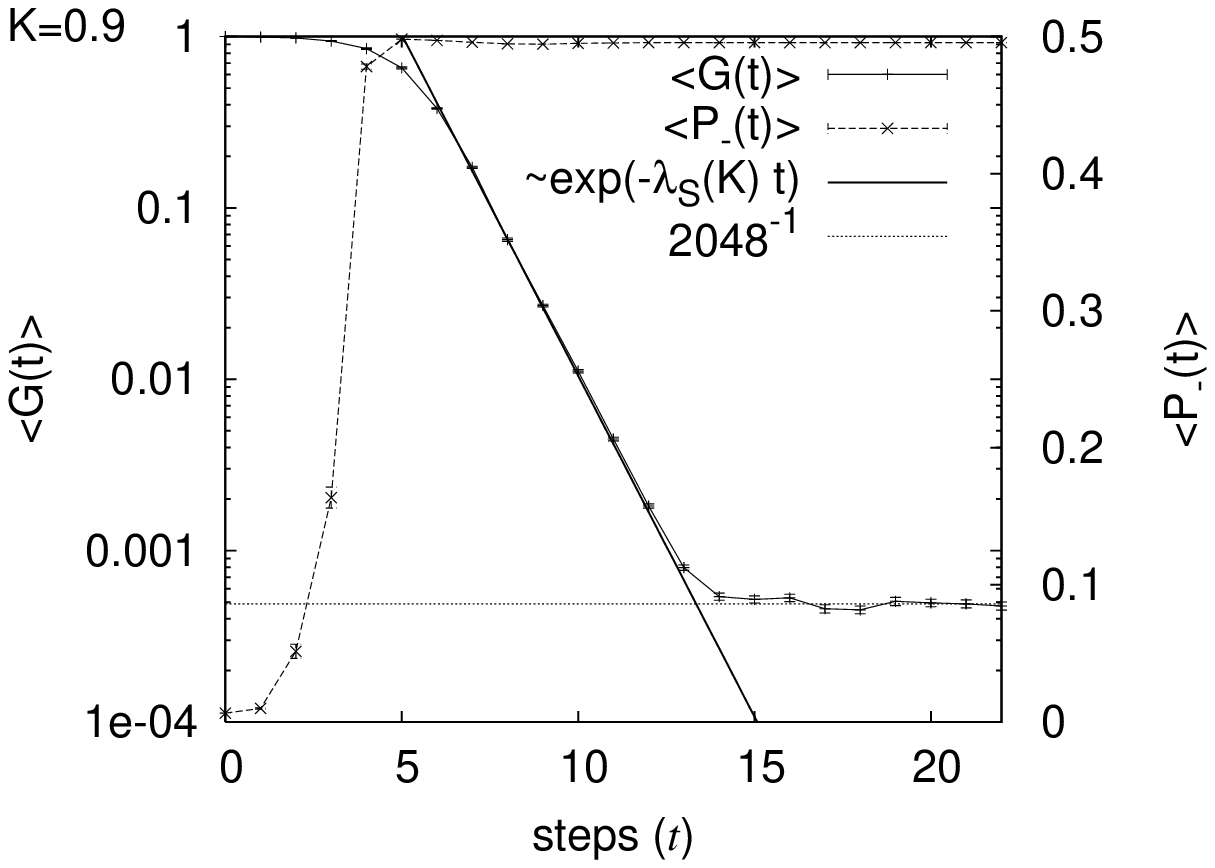}
\includegraphics[width=7cm]{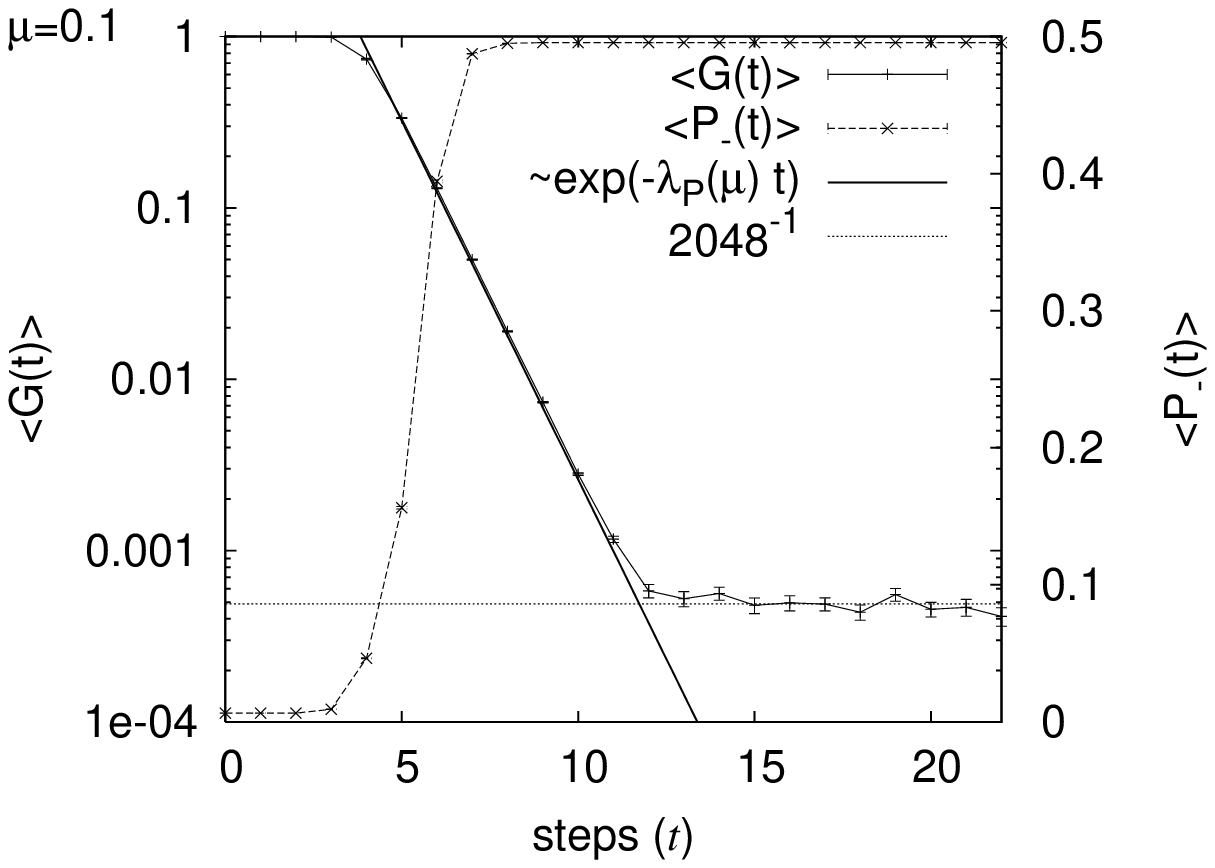}
\hbox to 16cm{\hfil(a)\hfil(b)\hfil}
\caption{Time evolution of the average relative QCF $\ave{G(t)}$ and of
the average portion of negative WF $\ave{P_{-}(t)}$ at $N$ = 2048 in the
Sawtooth map at $K=0.9$ (a) and Perturbed cat map at $\mu=0.1$ (b). The averages are taken over 500 initial Gaussian packets.}
\label{pic:wf_neg}
\end{figure}

Another interesting aspect of dynamics on the torus is that we can clearly see the connection between decay of QCF and the negativity of Wigner functions. In the figure \ref{pic:wf_neg} we show some numerical results where we plot together the relative QCF $G(t)$ and the relative fraction of phase space $P_{-}(t)$ supporting the Wigner function with the negative value, starting from the same initial Gaussian packets. From a detailed numerical study of statistical properties of the Wigner function for chaotic maps on compact phase spaces \cite{horvat03} we know that $P_{-}(t)$ should saturate in time to the value predicted by the {\em random wave model}
\begin{equation}
  P_{-,{\rm rand}}
  = \frac{1}{2} - \frac{1}{\sqrt{2\pi}}\int_0^r e^{-x^2/2} \dd x
  = \frac{1}{2} - \frac{r}{\sqrt{2\pi}} + O(r^2)
\label{eq:P_plateau}
\end{equation}
where $r = \mean{W}/\sigma_w$ is the ratio between the average Wigner function $\mean{W}$ and its standard deviation $\sigma_w$. For the considered Wigner function formulations on the torus we have
\begin{equation}
  {\rm continuous:}~r = (N-1)^{-1/2}\>,\qquad
  {\rm discrete:}~r = (4N-1)^{-1/2}\>.
\label{eq:p_ratio}
\end{equation}
We see that decrease of QCF closely follows the increase of $P_{-}(t)$, but there is obviously a strong difference between the discontinuous piece-wise linear (i.e. Sawtooth map) and non-linear smooth maps (i.e. Perturbed cat map).
Roughly speaking, the QCF decay appears when the Wigner function starts to develop, due to genuine quantum interference effects, Gaussian statistics for its values distribution. This correspondence is sharp for linear maps with discontinuities, like the saw-tooth map, whereas for nonlinear continuous maps, like perturbed cat map, the relaxation of Wigner function statistics happens little later than the time scale $T_1$ of QCF decay. Note that such behaviour has been observed for several other, different values of $N$ and as well as different values of parameters of the two representative models, not shown in the figure.

\section{Conclusion}

In the present paper we have discussed global phase space aspects of quantum-classical correspondence. We have defined a quantity, called quantum-classical fidelity (QCF), which characterizes $L^2$ norm distance between time evolving classical and quantum phase space distributions, starting from the same initial function, e.g. Gaussian. Quantum phase space distributions are described using Wigner-Weyl formalism.\par
Using heuristic arguments and numerical simulations, we conclude that QCF is characterized by two Ehrenfest-type time scales for classically chaotic dynamics and effectively compact phase spaces. QCF remains close to its maximal value up to time $T_1$, and after that starts to decay exponentially until it saturates to a plateau given by finite effective Hilbert space dimension, at time $T_2$. Both time scales are proportional to logarithm of effective Planck constant. Our results strongly suggest that the rate of exponential decay of QCF, after $T_1$, is given by the maximal Lyapunov exponent of the corresponding classical dynamics.\par
We have given heuristic arguments for our results in the case of Euclidean phase space geometry, and in addition, we have provided numerical results for generic chaotic maps on compact toroidal phase space. We outline subtle but significant differences between linear discontinuous maps and non-linear maps.\par
We have suggested that understanding of QCF may provide a key to
the correspondence between quantum and classical Loschmidt echoes,
namely we have proven an inequality which shows that quantum
Loschmidt echo simply follows the classical one up to time scale
$T_1$.\par
Several open problems and directions of future work can be suggested. Rigorous work is needed to understand the above results. One possible direction is the perturbative treatment of the Egorov property, like for example in the perturbed cat map studied above. Further, one may investigate the behavior of QCF for integrable systems and (weakly) chaotic systems with more complicated classical phase space structure, for example of non-uniformly hyperbolic or even KAM
type.

\section*{Acknowledgments}

 We are grateful to S.~Nonnenmacher for communicating to us his results \cite{Nonne} prior to publication, and for interesting discussions.
TP acknowledges support from the grants J1-7347 and P1-0044 of Slovenian 
Research Agency.

\section*{References}

\end{document}